# Assessing site-specific enhancements imparted by hyperpolarized water in folded and unfolded proteins by 2D HMQC NMR

Or Szekely[†,$], Gregory Lars Olsen[†,&], Mihajlo Novakovic[†], Rina Rosenzweig[‡], Lucio Frydman*[,†]

Departments of [†]Chemical and Biological Physics and [‡]Structural Biology, The Weizmann Institute of Science, 234 Herzl Street, Rehovot 760001, Israel

## Abstract

Hyperpolarized water can be a valuable aid in protein NMR, leading to amide group $^1$H polarizations that are orders of magnitude larger than their thermal counterparts. Suitable procedures can exploit this to deliver 2D $^1$H-$^{15}$N correlations with good resolution and enhanced sensitivity. These enhancements depend on the exchange rates between the amides and the water, thereby yielding diagnostic information about solvent accessibility. This study applied this "HyperW" method to four proteins exhibiting a gamut of exchange behaviors: PhoA$^{(350\text{-}471)}$, an unfolded 122-residue fragment; barstar, a fully folded ribonuclease inhibitor; R17, a 13.3 kDa system possessing folded and unfolded forms under slow interconversion; and drkN SH3, a protein domain whose folded and unfolded forms interchange rapidly and with temperature-dependent population ratios. For PhoA4$^{(350\text{-}471)}$ HyperW sensitivity enhancements were ≥300×, as expected for an unfolded protein sequence. Though fully folded barstar also exhibited substantially enhancements; these, however, were not uniform, and according to CLEANEX experiments reflected the solvent-exposed residues. R17 showed the expected superposition of ≥100-fold enhancements for its unfolded form, coexisting with more modest folded counterparts. Unexpected, however, was the behavior of drkN SH3, for which HyperW enhanced the unfolded but even more certain *folded* protein sites. These preferential enhancements were repeatedly and reproducibly observed; a number of explanations – including three-site exchange magnetization transfers between water, unfolded and folded states; cross-correlated relaxation processes from hyperpolarized "structural" waters and labile sidechain protons; and the possibility that faster solvent exchange rates characterize certain folded sites over their unfolded counterparts– are considered to account for them.

## Introduction

Nuclear magnetic resonance (NMR) plays an irreplaceable role in biophysical studies. NMR can tackle complex systems such as proteins in solution, under native or near-physiological conditions, and provide information about the structures and dynamics of these systems with atomic resolution. Despite this potential, NMR in general –and NMR of large biomolecules in particular– suffers from inherent sensitivity issues. Improving sensitivity and signal-to-noise ratio in NMR has therefore been the focus of extensive efforts, including the use of hyperpolarization methods that can impart orders-of-magnitude sensitivity enhancements to a variety of solutions and solids.[1-5] Out of all methods for nuclear hyperpolarization, dissolution DNP stands out in its generality to enhance the sensitivity of high-field solution-state NMR and MRI measurements.[6-9] However, the *ex situ* nature of this approach –where the sample is hyperpolarized in one magnet under cryogenic conditions and then transferred as a liquid to another system for its eventual observation– fails when attempted on large biomolecules subject to very fast low-field relaxation processes. Hyperpolarized water[10-12] (HyperW) NMR was recently introduced to overcome this limitation, and enable the study of proteins and nucleic acids.[13] HyperW NMR relies on the fact that $H_2O$'s protons can be hyperpolarized by dissolution DNP into the tens of percent, and if suitably handled their relaxation times can reach into the tens of seconds. These protons, being labile, can then spontaneously exchange with groups in biomolecules –for instance with amides in unfolded proteins or intrinsically disordered proteins/domains (IDPs/IDDs). This will then hyperpolarize the amide protons for long enough to enable the acquisition of 2D $^1$H-$^{15}$N NMR correlations, particularly if the direct excitation of the water reservoir that is constantly supplying the amides with polarization, is avoided. Initial HyperW biomolecular 1D and 2D NMR experiments delivered considerable sensitivity enhancements –≥300× over their thermal counterparts– for mixtures of short peptides,[14] albeit with poor spectral resolution. Hyperpolarized water also enabled the study of weak protein interactions[15] and IDPs.[16] More recently,[17] this method was used to achieve substantial enhancements for the Parkinson's-disease-associated IDP α-synuclein as well as full 2D $^1$H-$^{15}$N HMQC NMR resolution using an optimized water-injection experimental setup. With sensitivity enhancement values resolved for each amide group in the polypeptide, a simple model based on the Bloch-McConnell equations was then developed that could translate these HyperW enhancements in terms of the residue-specific dynamics characterizing amide/water exchanges for α-synuclein.

This work extends these optimized HyperW measurements to a wider variety of protein structures. These included the fully unfolded protein fragment PhoA$^{(350\text{-}471)}$ stemming from Alkaline

Phosphatase, for which the sensitivity enhancements observed were substantial and distributed randomly throughout the sequence. Also included was barstar, a protein which, although fully folded, also evidenced double-digit enhancements for certain residues —particularly solvent-exposed ones, for which ancillary CLEANEX experiments confirmed that the HyperW method acts as a kind of "exchange filter". The third kind of system analyzed involved equilibria between coexisting unfolded and folded conformations, interconverting at different rates: these included the R17 domain of chicken-brain α-spectrin, and a terminal Src homology 3 domain from Drosophila, drkN SH3. In both cases the HyperW approach was able to light up both coexisting folded/unfolded populations, and to deliver from these enhancements site- and state-discriminated pictures of solvent accessibility for both folded and unfolded forms. For the R17 dimer these were, as expected, ca. 5-10× higher for the unfolded form. Different, however, was the behavior observed for drkN SH3 –where repeated experiments consistently indicated equal or larger enhancements for residues in the *folded* form, than in their unfolded counterparts. Potential mechanisms and consequences of such observations, which depart from a paradigm whereby hyperpolarized water acts as a simple reporter of solvent accessibility, are assessed.

## Materials and methods

**Sample preparation.** An *E. coli* PhoA (residues 350-471) fragment was produced and purified as described by Saio *et al.*[18] This $PhoA^{(350\text{-}471)}$ (PhoA4) was cloned into a pET-16b vector. The final gene incorporates an N-terminal Hexa-His-MBP tag followed by a Tobacco Etch Virus (TEV) protease cleavage site. A culture of BL21(DE3) harboring the PhoA4 plasmid was grown in M9 minimal medium supplemented with 1 g/L $^{15}$N-labeled ammonium chloride and Ampicillin (100 µg/mL). The culture was induced at $OD_{600}$ 0.5 and overexpressed at 18 °C overnight. The protein was isolated from the lysate using a Ni-NTA column and the His tag was removed by incubation with TEV protease overnight at 4 °C. PhoA4 was separated from the tag and TEV protease by passing it over a Ni-NTA column and further purified on a Superdex 75 size exclusion column (GE Healthcare). The samples containing PhoA4 were buffer exchanged to a concentration of 1.5 mM or 0.35 mM in a 99.9% $D_2O$ buffer (50 mM HEPES, pD 7.5, 50 mM KCl). For HyperW dissolutions, 140-150 µL aliquots of this solution were placed in a 5 mm NMR tube for their subsequent analysis. Following the hyperpolarized water injection, the sample was thus diluted to either 0.6 mM or 0.13 mM protein. For the reference, high protonated water-content sample, 35 µL of the 0.35 mM PhoA4 solution was diluted with a 90% $H_2O$ buffer (50 mM HEPES, pH 7.5, 50 mM KCl) to a concentration of 0.13 mM protein in 82.5% $H_2O$.

Barstar was produced and purified as described by Schreiber *et al.*[19] In brief, a culture of BL21(DE3)pLysS harboring a plasmid encoding a mutated barstar (C40A, C82A) was grown in M9 minimal medium supplemented with 1 g/L $^{15}N$-labeled ammonium chloride, Ampicillin (100 µg/mL) and Chloramphenicol (17 µg/mL). The culture was induced at $OD_{600}$ 0.6 with 200 µM isopropyl β-D-1-thiogalactopyranoside (IPTG) and grown overnight at 30 °C. The cell pellet was resuspended in buffer (10 mM Tris pH 8, 1 mM EDTA, 100 mM NaCl, 1 mM PMSF, 50 mg/ml Lysozyme and DNAse) and disrupted with a cooled cell disrupter (Constant Systems) followed by centrifugation. Barstar found in the soluble fraction was isolated by precipitation with 40-80% ammonium sulfate. After centrifugation the pellet was resuspended in a minimal volume of buffer (50 mM Tris pH 8, 100 mM NaCl), injected to a gel filtration column (Hiload_Superdex_75_26/60, GE Healthcare), and pre-equilibrated with the same buffer. Final purification on an anion exchange column (HiTrap_Q_HP, GE Healthcare) involved elution with 300 mM NaCl. The fractions containing barstar were dialyzed to DDW and lyophilized. For the HyperW experiments, lyophilized barstar was reconstituted in a $D_2O$ buffer (50 mM Sodium Phosphate, pD 7) at a concentration of ~ 4 mM. 130-140 µL aliquots of this solution were placed in a 5 mm NMR tube for their subsequent analysis. Following the hyperpolarized water injection, the sample was thus diluted to 1.3-1.6 mM protein. For the reference, high protonated-water-content sample, a post-injection sample was lyophilized to dryness and subsequently reconstituted in the same volume of a 90% $H_2O$ buffer (50 mM Sodium Phosphate, pH 7), to give rise to the same final protein concentrations of 1.6 mM.

The R17 domain dimer was produced and purified as described by Sekhar *et al.*[20] In brief, the gene encoding L90A R17 domain of chicken-brain α-spectrin was cloned into a pET-29b(+) vector. The final gene incorporates an N-terminal hexa-His tag followed by a short linker and a TEV protease cleavage site. A culture of BL21(DE3) cells harboring the R17 plasmid was grown at 37 °C in M9 minimal medium supplemented with 1 g/L $^{15}N$-labeled ammonium chloride and Kanamycin (50 µg/mL). The culture was grown to $OD_{600}$ 0.8 and overexpressed at 22 °C overnight. The protein was isolated from the lysate using a Ni-NTA column and the His tag was removed by incubation with TEV protease overnight at 4 °C. R17 was separated from the tag and TEV protease by passing it over a Ni-NTA column and further purified on a Superdex 75 size exclusion column (GE Healthcare). The protein eluted as two peaks (monomer and dimer), and the dimer fractions were collected. The samples containing $^{15}N$ labelled R17 dimer were buffer exchanged to a concentration of 1.23 mM in a 99.9% $D_2O$ buffer (50 mM HEPES, pD 7.5, 50 mM KCl). For the HyperW dissolution experiment, a 140 µL aliquot of this solution was placed in a 5 mm NMR tube for its subsequent analysis. Following the hyperpolarized water injection, the sample was thus diluted to a protein concentration of 0.57 mM.

The drkN SH3 domain was produced and purified as described by Sekhar *et al.*[20] The gene for the SH3 domain of *Drosophila melanogaster* Enhancer of sevenless 2B protein (drkN SH3) was cloned into the pET-28 vector using PCR amplification (Kapa Hifi, Kapa Biosystems, MA, U.S.A.) followed by Gibson assembly (New England Biolabs, MA, U.S.A.). The final gene incorporates an N-terminal Hexa-His tag followed by a TEV protease cleavage site. A culture of BL21(DE3) cells harboring the drkN SH3 plasmid was grown at 37 °C in M9 minimal medium supplemented with 1 g/L $^{15}$N-labeled ammonium chloride and Kanamycin (50 mg/L). The culture was grown to $OD_{600}$ 0.8 and overexpressed at 25 °C overnight. The protein was isolated from the lysate using a Ni-NTA column under denaturing (6 M guanidinium chloride) conditions. The unfolded protein was refolded on the column before elution by lowering the denaturant concentration stepwise from 6 to 4, 2, 1 and finally to 0 M. The His tag was removed by incubation with TEV protease overnight at 4 °C. DrkN SH3 was separated from the tag and TEV protease by passing it over a Ni-NTA column and further purified on a Superdex 75 size exclusion column (GE Healthcare). The samples containing drkN SH3 were buffer exchanged to concentrations of 0.8 mM or 1.3 mM in a 99.9% $D_2O$ buffer (50 mM HEPES, pD 7.5, 50 mM KCl). For HyperW dissolutions at 50 °C, 130 or 80 µL aliquots of the 0.8 mM solution, or 140 µL of the 1.3 mM solution, were placed in a 5 mm NMR tube for their subsequent analysis. For HyperW dissolutions at 37 °C, 145 µL of the 1.3 mM solution were placed in the 5 mm NMR tube. Following four hyperpolarized water injections, the sample was thus diluted to 0.26 / 0.16 / 0.59 mM protein (at 50 °C), or 0.51 mM (at 37 °C). For the reference, high protonated-water-content samples, the first two post-injection samples were lyophilized to dryness and subsequently reconstituted in the same volume of a 90% $H_2O$ buffer (50 mM HEPES, pH 7.5, 50 mM KCl), to give rise to the same final protein concentrations of 0.26 / 0.16 mM. The third high-water-content sample was prepared by dilution of 145 µL of 1.3 mM in a 99.9% $D_2O$ buffer with a 100% $H_2O$ buffer (50 mM HEPES, pH 7.5, 50 mM KCl to a concentration of 0.52 mM protein and 87.4% $H_2O$. The latter was used as a reference both at 50 °C and 37 °C. Further sample preparation details are given in the figure captions.

**Dynamic Nuclear Polarization.** Water was hyperpolarized using an Oxford Instrument Hypersense® equipped with a 3.35 T magnet. The system was modified by adding to the Oxford-supplied E2M80 vacuum pump, an EH-500 Edwards booster capable of taking the operating pressure to 1 torr. Polarization was thus typically done at ~1.05-1.30 K. DNP was achieved by irradiating at ~94.1 GHz a 10 mM 4-amino-TEMPO (4AT) nitroxide radical, dissolved in ca. 100 µL solutions containing 15% glycerol and 85% $H_2O$ (v/v). Optimized microwave power levels and pumping time were 100 mW (nominal) and 180 min. Following this irradiation, samples were dissolved with a 99.9% $D_2O$ buffer; approximately 300 µL of the melted, hyperpolarized samples were then transferred into the NMR magnet using a pre-heated (60 °C) tubing line, and injected into a 5 mm tube containing the targeted biomolecules dissolved in buffered $D_2O$.

**Injection Setup.** Sample injections were carried out on an automated pressurized system achieving robust, reproducible transfers, with minimum bubble formation. The system and its design have been described elsewhere.[14,17] In brief, it relies on a two-state valve operation,[21-23] controlling the filling of the NMR tube using a three-port accessory involving both forward and backward gas pressures and controlled by an Arduino®-based software.[23] Following previous optimization of the injection setup for obtaining high resolution two-dimensional (2D) protein spectra[17], the injection system driving pressure was set to a gradient between 17 and 3.5 bar.

**NMR spectroscopy.** Post-dissolution NMR experiments were conducted using a 5 mm liquid-nitrogen-cooled "Prodigy®" probe in a 14.1 T Bruker magnet interfaced to an Avance III® console. These experiments included 2D NMR acquisitions, which were triggered upon injecting the hyperpolarized water sample into the NMR tubes waiting with their samples inside the magnet bore. Experiments were carried out at nominal temperatures of either 37 or 50 °C, as detailed below. In view of the claims made below for the case of co-existing folded/unfolded protein states, particular attention was paid to the thermal reliability and uniformity of the sample temperatures, resulting upon co-mixing the pre-heated hyperpolarized water with the pre-heated protein solution waiting inside the NMR tube. An idea of the thermal gradients and thermal stabilization of the ensuing mix is presented in Supporting Figure S1, which analyzes the stabilization of the NMR signal throughout a 2D HyperW NMR acquisition performed at 50 °C, on the basis of two water-enhanced residues with temperature-sensitive resonances. It follows from this analysis that the temperature stabilizes to within one degree of the target value, within ≈10 s within the acquisition. 2D HyperW NMR spectra were acquired using the $^{1}H$-$^{15}N$ HMQC sequence given in Supporting Figure S2.[14,17] This sequence fully excites and echoes the downfield amide region selectively and employs minimal recycle delays,[24-25] in order to maximize the signal from the hyperpolarized exchangeable sites while minimizing the water depolarization losses. Unless otherwise noted, thermal equilibrium measurements were carried out on the same sample with the same hardware and using the same pulse sequence but with longer recycle delays, to obtain reliable measures of the HyperW site-specific enhancements. Ancillary CLEANEX-PM[26] experiments were collected on the same spectrometer and probe at 50 °C or 37 °C. ZZ-exchange and methyl-TROSY experiments were measured on 5 mm cryogenically-cooled probes in 14.1 T or 18.8 T Bruker magnets interfaced to Bruker AvanceNeo or AvanceIII® consoles respectively, at 50 °C or 37 °C. All NMR data were processed using the Bruker® Topspin® software and subsequently plotted and analyzed using Matlab®. Non-uniformly sampling (NUS) using a Poisson-gap sampling schedule and spectral reconstructions was implemented using the hmsIST software,[27] in combination with Topspin®.

## Results and Discussion

### *HyperW on a disordered peptide: The Alkaline Phosphatase 350-471 fragment PhoA4*

Disordered proteins are natural candidates for water-based hyperpolarization enhancements, since their amide protons are exposed to the solvent. The ensuing rapid amide/water exchange rates, should facilitate substantial enhancements according to HyperW's Bloch-McConnell model.[17,28] An example of this is provided by the fully disordered protein fragment PhoA$^{(350\text{-}471)}$ (PhoA4). This 122 residue polypeptide is completely unfolded under reducing conditions.[18,29-31] Consistent with this, the NMR chemical shifts of the PhoA4 fragment match the values known for the same residues in the full length protein.[18] Figure 1 compares a representative 2D $^{1}$H-$^{15}$N HMQC spectrum measured at 50 °C for this unfolded $^{15}$N-labeled protein upon injection of hyperpolarized water, against a thermal counterpart, both containing only ca. 2% protonated $H_2O$. Notice that in this conventional spectrum, measured using the same sample at the same temperature, most peaks broaden beyond detection due to fast exchanges with the solvent. While this exposure conspires against normal 2D NMR, it facilitates the magnetization transfer from the hyperpolarized water, leading to strongly enhanced peaks. This evidences a certain complementarity between HyperW-based and conventional HMQC acquisitions. While enhancements can be calculated only with large errors when the hyperpolarized spectrum is compared against a thermal spectrum collected from the dissolution DNP sample, peaks emerge from the noise if the PhoA4 HMQC spectrum is measured with the same sequence in a fully protonated $H_2O$ buffer at 50 °C (Supporting Fig. S3). The average sensitivity enhancement that can be then calculated for the unfolded PhoA4 fragment is ~260× when considering all peaks in the spectrum. This high enhancement is typical of what we have obtained in unfolded protein injections, using our 14.1 T NMR and hyperpolarization setup.

By comparing to the BMRB entry of the full length PhoA[32] and extrapolating according to the changes that peaks undergo with temperature and pH, several peaks in the HyperW spectrum (Fig. 1) can be tentatively assigned. With these assignments, enhancements can be calculated for specific residues; the average enhancement for these resolved residues (Fig. 2) is ~130×, substantially lower than what arises by considering the overall peak volume of the spectra. It is also clear that within this assignable set there are sites which get enhanced much more than others, a heterogeneity that could reflect water accessibility and/or local residue charges. To evaluate the influence of the former we relied on secondary structure propensity (SSP) scores, which can range from +1 for a completely structured α-helix, to 0 in a disordered structure, to -1 for a β-sheet.[33] Saio et al[18] calculated SSPs for this protein fragment; the grey bars in Fig. 2 illustrate these parameters as a function of the primary sequence. Also added to Fig. 2 are orange and red squares indicating positively and negatively charged residues, respectively. Unlike what had been previously observed for α-synuclein, the sensitivity enhancements evidenced by HyperW HMQC do not appear to correlate with these electrostatic charges in the sequence; the correlation arising between the enhancements and the SSP values is also questionable –if present at all (Supporting Figure S4).

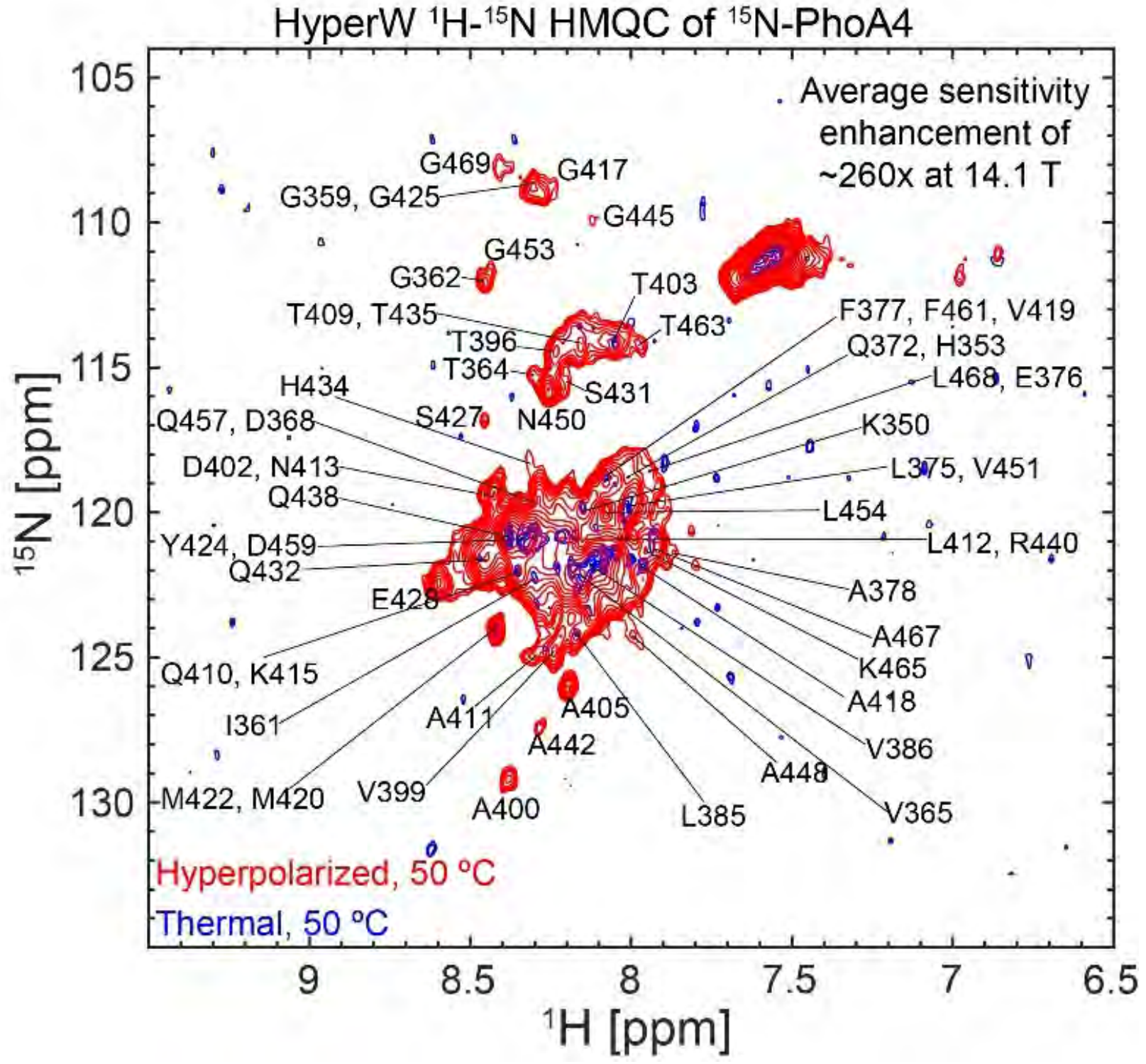

**Figure 1.** Comparison between 2D HyperW (red) and conventional (blue) $^{1}H$-$^{15}N$ HMQC spectra measured on $^{15}N$-PhoA4. 2.8 mL of super-heated buffered $D_2O$ (50 mM HEPES, pD 7.5, 50 mM KCl) was used to dissolve an 85/15 water/glycerol pellet containing 10 mM 4-amino-TEMPO. This pellet had been polarized at 1.12 K for ~ 3 hrs 30 min using 100 mW of microwave irradiation at 94.195 GHz. ~240 μL of the resulting hyperpolarized water solutions were injected into a 5 mm NMR tube containing ~140 μL of a 1.5 mM $^{15}N$-labeled PhoA4 solution. Partial tentative assignment of residues indicated by single-letter amino acid codes is done based on the BMRB entry of the full length PhoA.[32] Both spectra were recorded at 50 °C using 64 hypercomplex $t_1$ increments and hypercomplex[34] acquisition covering indirect- and direct-domain spectral widths of 6009.6 and 1825.8 Hz. The HyperW spectrum was recorded using two phase-cycled scans per $t_1$. Additional experimental parameters: 14.1 T Prodigy®-equipped NMR; total acquisition times of 73 s for the hyperpolarized spectrum (acquisition time of 213.0 ms, repetition delay of 0.037 s) and 14 hrs 12 min for the thermal spectrum (320 scans per $t_1$ increment, acquisition time of 213.0 ms, and a repetition delay of 1 s).

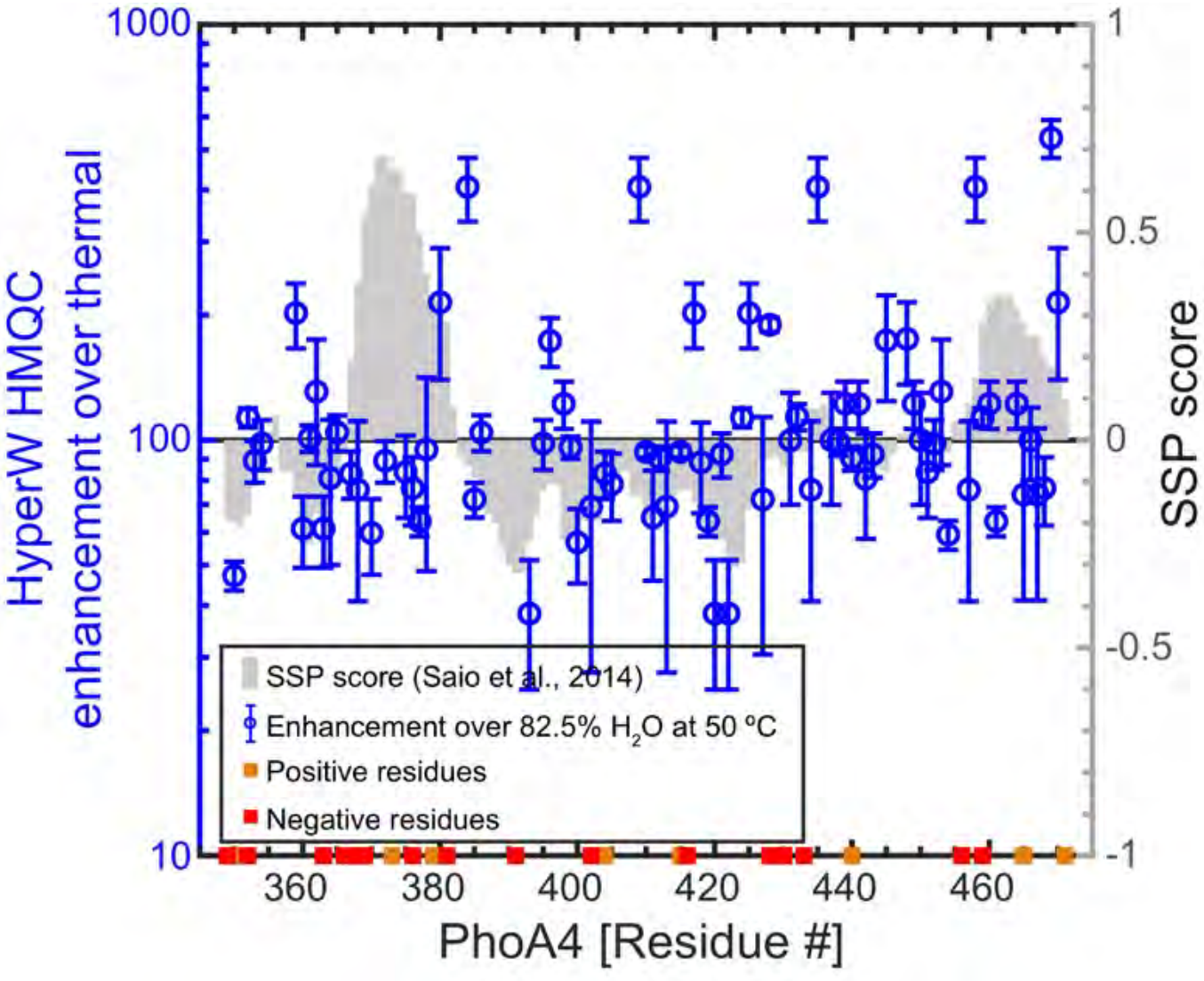

**Figure 2.** HyperW HMQC sensitivity enhancements calculated for resolved residues in the $^{15}N$-labeled PhoA4 protein fragment. The sensitivity enhancements were extracted by comparing peak volumes between the HyperW HMQC spectrum (e.g., Fig. 1, red) and thermal equilibrium spectra measured in an 82.5% $H_2O$ buffer (Supporting Fig. S3). The values are averaged for three HyperW HMQC experiments, after normalizing to the $H_2O$ proton enhancement in each experiment; the "error bars" reflect the scattering obtained over the course of these repeated injections for each residue. Sensitivity enhancements compared against SSP scores (grey bars) given in the literature[18] based on NMR $^{13}C_\alpha$ and $^{13}C_\beta$ chemical shifts. Charged residues are also mapped on the sequence with orange (positively charged) and red (negatively charged) squares.

### *HyperW NMR on a fully structured peptide: Barstar*

Barstar is an 89 residue protein from *Bacillus amyloliquefaciens* bacteria with a well-defined, folded structure.[35-36] Extensive work has been done on this protein as a model of folding,[37-42] with most crystallographic and folding studies centering on the C40/82A mutant. We thus chose this well-studied construct to test the outcome of HyperW HMQC experiments on a well-folded paradigm.

Figure 3 shows the sensitivity enhancements that HyperW HMQC NMR at 50 °C and 2% $H_2O$, imparts on this double C40/C82A barstar mutant (note that the protein is stable at this temperature, as its $T_m$ is ≈70-75 °C at the pH~7–8 used in this study[43-45]). The good resolution delivered by the post-DNP rapid injection system provides clearly resolved resonances with chemical shifts that are characteristic of well-folded structures; this is in agreement to what has been recently reported by Kadeřávek et al on the folded protein ubiquitin,[46] regarding the compatibility of water-derived hyperpolarization with studies of folded biopolymers. In fact, after taking into account the changes in chemical shifts with temperature, it was possible to assign most of the peaks in the HyperW HMQC (80 out of 89) based on literature data;[47] these are annotated in Figure 3A. Despite this site resolution it is also clear that peaks along the indirect dimension of the HyperW experiments, are broader than their thermally-collected counterparts. This results from the limited lifetime of the water hyperpolarization, which, driven by $T_1$, by chemical exchange with the biomolecule, and by decays induced by pulse nonidealities (pulses are tuned to minimally touch the water resonance), put an upper bound on the number of points that can be conventionally sampled along the $t_1$ domain. For the kind of systems hereby analyzed, ca. 30 to 60 sec is the time available for probing the indirect dimension of the 2D NMR spectra. Non-uniform sampling (NUS)[48-49] should be able improve this resolution further while retaining the same overall experimental time. Figure 3B illustrates this with HyperW and thermal spectra recorded and processed on the same sample with NUS, where the effective $t_1$ evolution time was increased four-fold but the actual sampling covered only a 25% fraction of a conventional acquisition. The improvement in resolution along the indirect dimension for both experiments (thermal and hyperpolarized) is evident. Overall the average sensitivity enhancements in both regularly and non-uniformly sampled experiments are comparable, as the longer evolution times employed in the latter are offset by the smaller number of points (and hence fewer pulses) employed.

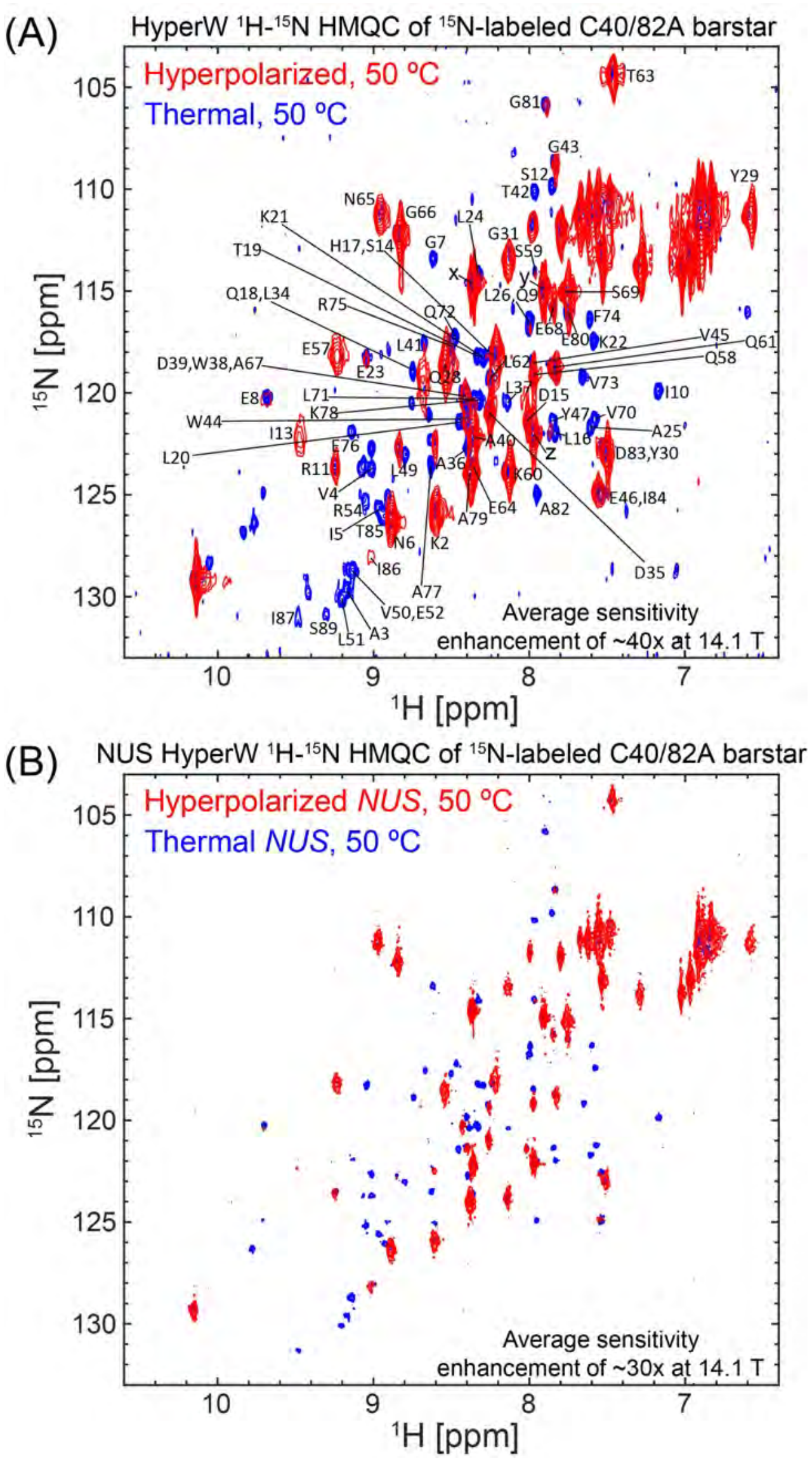

**Figure 3.** (A) Comparison between 2D HyperW (red) and conventional (blue) $^1$H-$^{15}$N HMQC spectra measured on $^{15}$N-labeled barstar C40/82A double mutant. 2.8 mL of super-heated buffered $D_2O$ (50 mM sodium phosphate, pD 7) was used to dissolve an 85/15 water/glycerol pellet containing 10 mM 4-amino TEMPO. The pellet was polarized at 1.20 K for ~ 3 hrs using 100 mW nominal microwave irradiation at 94.195 GHz. ~215 µL of the resulting hyperpolarized water solutions were injected into a 5 mm NMR tube containing ~140 µL of a ~4 mM $^{15}$N-labeled barstar mutant solution. Partial assignment of 80 (out of 89) residues, indicated here by single-letter amino acid code on the basis of Wong et al.[47]. The three peaks marked x, y and z are unassigned and are attributed to free amino acids in the sample. Both spectra were recorded at 50 °C using 64 hypercomplex $t_1$ increments[34] covering indirect- and direct-domain spectral widths of 7211.5 and 1825.8 Hz. The HyperW spectrum was recorded using two phase-cycled scans per $t_1$. Total experimental time of 72 s for the hyperpolarized spectrum (acquisition time of 213.0 ms, repetition delay of 0.037 s) and 2 hrs 51 min for the thermal spectrum (64 scans and an acquisition time of 213.0 ms per $t_1$ increment, repetition delay of 1 s). (B) Non-uniform sampling improves HyperW resolution. Non-uniformly sampled 2D HyperW (red) and non-uniformly sampled conventional (blue) $^1$H-$^{15}$N HMQC spectra were measured on the $^{15}$N-labeled barstar C40/82A double mutant. 2.8 mL of super-heated buffered $D_2O$ (50 mM sodium phosphate, pD 7) was used to dissolve the 85/15 water/glycerol pellet containing 10 mM 4-amino TEMPO. The pellet was polarized at ~1.18 K for ~ 3 hrs 03 min using microwave irradiation of 100 mW, 94.195 GHz. ~250 µL of the resulting hyperpolarized water solutions were injected into a 5 mm NMR tube containing ~130 µL of a ~4 mM $^{15}$N-labeled barstar mutant solution. Both spectra were recorded at 50 °C sampling 25% of 256 hypercomplex $t_1$ increments[34] covering indirect- and direct-domain spectral widths of 7211.5 and 1825.8 Hz, leading to a four-fold increase in maximum effective $t_1$ evolution. The HyperW spectrum was recorded using two phase-cycled scans per $t_1$. Total experiment times were ~80 s for the hyperpolarized spectrum (acquisition time of 213.0 ms, repetition delay of 0.037 s) and 11 hrs 50 min for the thermal spectrum (256 scans recorded and 213.0 ms acquisition time per $t_1$ increment, with a repetition delay of 1 s).

Identification of the individual peaks reveals a remarkably heterogeneous picture for the HyperW enhancements characterizing barstar, which range from <1× for some residues to >300× for others (Fig. 4). These sensitivity enhancements are calculated by comparing peak volumes between the HyperW HMQC spectrum (such as in Fig. 3A, red) and the thermal equilibrium spectrum measured for the same sample in 90% $H_2O$ buffer. In general, residues in loops and otherwise disordered regions of the folded conformation appeared enhanced to a greater extent than those in the structured areas, highlighting again the relation between HyperW signal increases and accessibility to the hyperpolarized solvent. However for other residues, including I13 and amides in helix-3 and helix-4 in the protein, the measured enhancements are also high. The close connection between these enhancements and water/amide exchange rates is further confirmed by CLEANEX-PM NMR,[26] an experiment designed to highlight water-exposed residues. In these experiments the water resonance is selectively excited and allowed to exchange over a variable mixing period $\tau_m$ with the amide proton spins. At the end of these mixing periods a fast HSQC sequence is used for detection, and the amide resonances peak intensities are monitored for every $\tau_m$ on a per-residue basis. Using short mixing times, only the fast-exchanging amides will have enough magnetization coming from water. The longer the amides protons are allowed to exchange with the water, the higher their magnetization will be. Figure 5A illustrates the close match between long-$\tau_m$ CLEANEX-PM experiments, and the HyperW data.

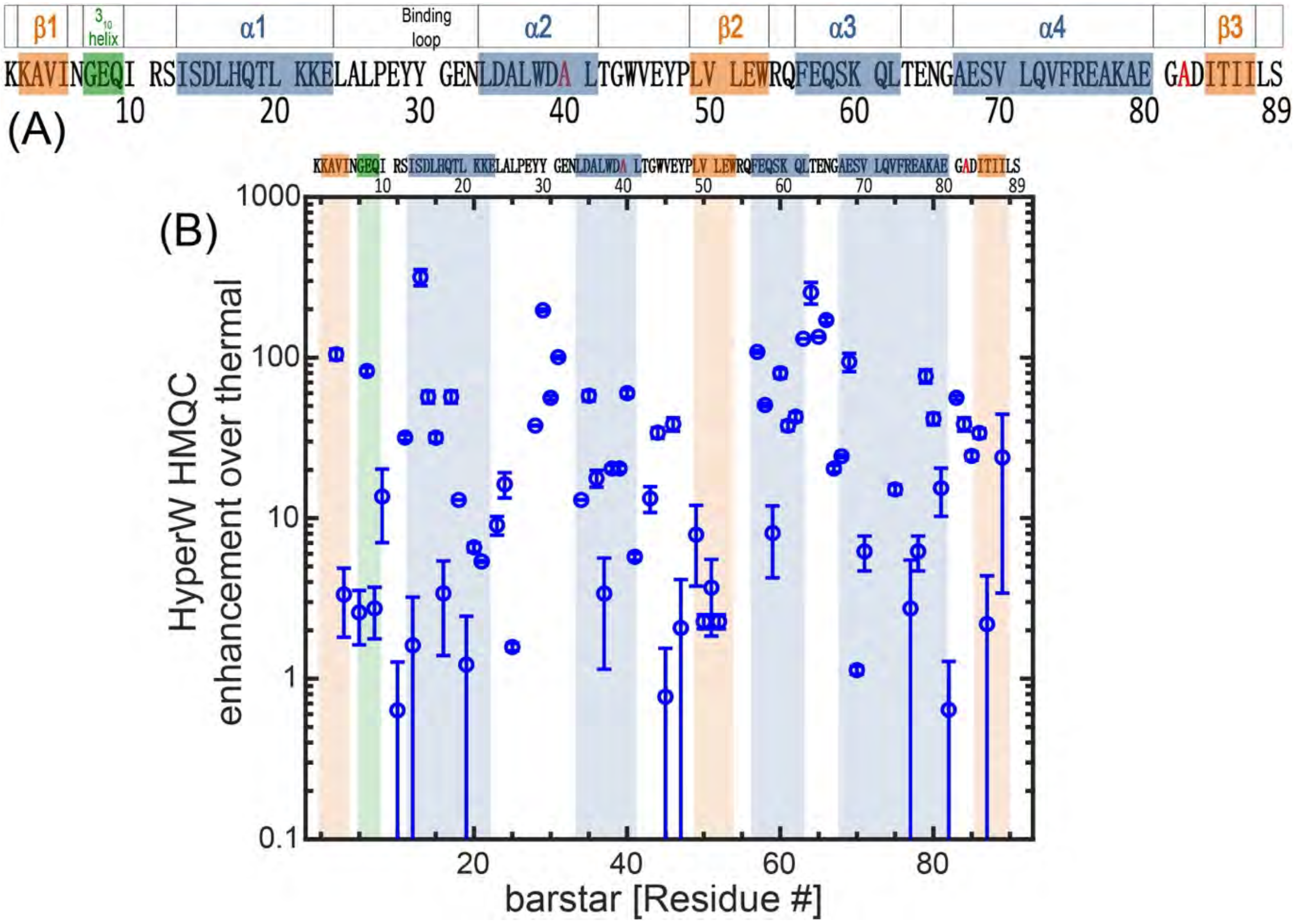

**Figure 4.** (A) 89-residue barstar C40/82A sequence analyzed in this study. Secondary structure elements[47] are denoted above the sequence and shaded in blue (α-helices), orange (β-strands) and green ($3_{10}$ helix). A flexible loop, which plays an important role in binding barnase[50-51] is also noted. The C40/82A mutations are shown in red. (B) HyperW HMQC sensitivity enhancements observed for the assigned residues of the $^{15}N$-labeled mutant. The sensitivity enhancements were calculated by comparing peak volumes between the HyperW HMQC spectrum (such as in Fig. 3A, red) and the thermal equilibrium spectrum measured for the same sample in 90% $H_2O$ buffer. The values are averaged for two HyperW HMQC experiments, after normalizing to the $H_2O$ proton enhancement in each experiment; "error bars" reflect the scattering of these experiments. Blue, orange and green shaded areas are drawn on the regions corresponding to the secondary structure elements in (A).

The theory for extracting exchange rates $k$ from CLEANEX is well established,[26,52-55] and is based on the equation

$$\frac{V}{V_0} = \frac{k}{R_{HN,app}+k-R_{H_2O,app}} \cdot \left[e^{-R_{H_2O,app}\cdot\tau_m} - e^{-(R_{HN,app}+k_{HN})\cdot\tau_m}\right] \quad (1)$$

where $V$ is the CLEANEX peak volume, $V_0$ the corresponding peak volume in a reference HSQC spectrum. $R_{HN,app}$ is the apparent relaxation rate for the amides, containing contributions from the longitudinal relaxation rate $1/T_1^{HN}$ and from the transverse relaxation rate $1/T_2^{HN}$, while the apparent relaxation rate for water is its longitudinal relaxation rate $R_{H2O,app} = 1/T_1^{W}$. The rate constant $k$ is related to the amide-water exchange rate $k_{HN}$ used in our previous analyses of HyperW signal enhancements[17] by $k_{HN} = X_{H2O}\cdot k$; since $X_{H2O}$ (the molar fraction of $H_2O$) ≈ 1, $k_{HN} \approx k$. CLEANEX-derived rates should thus be, within the uncertainty limits of the relaxation and overall DNP enhancement (ε) parameters, similar to those arising from HyperW methods. Figure 5B shows that there is indeed a relatively good correlation (r = 0.63, calculated in a linear enhancement vs $k_{HN}$ plot) between the measurements.

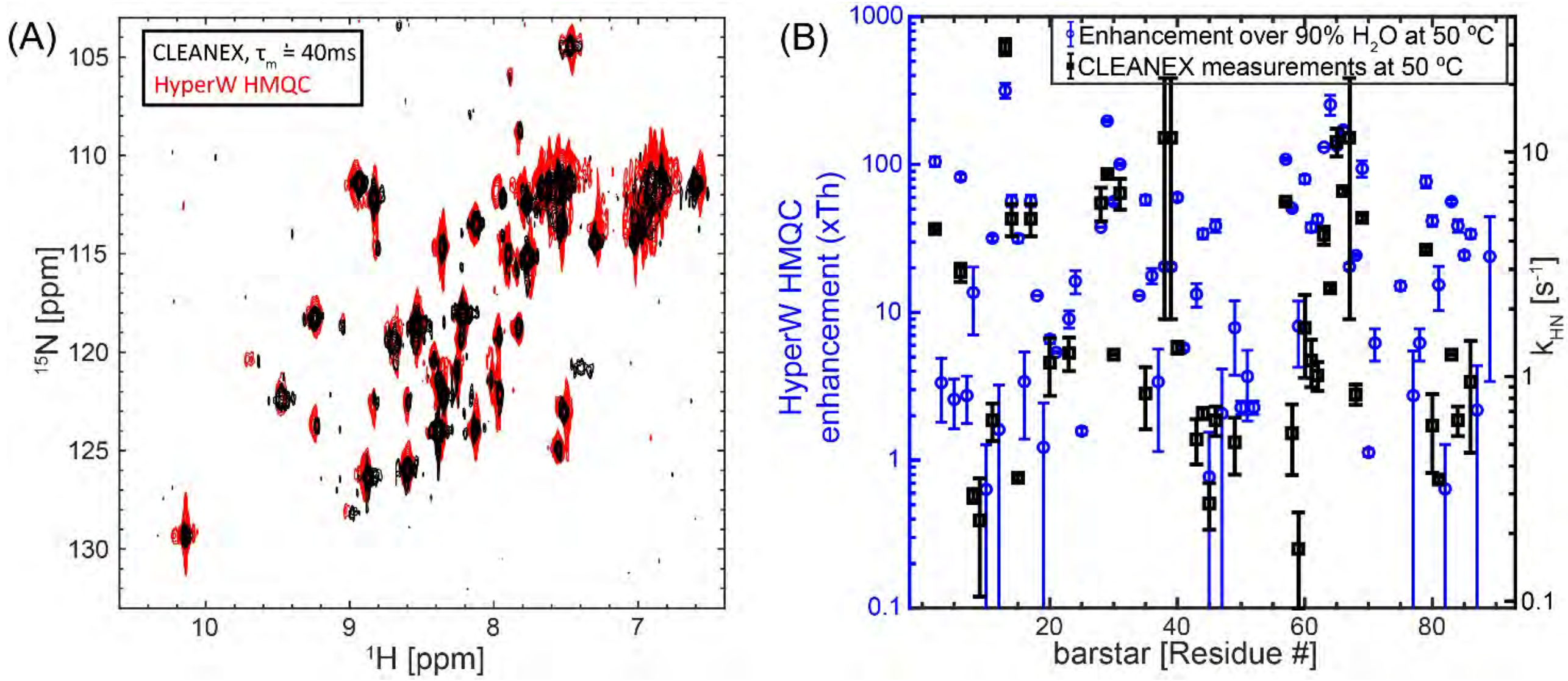

**Figure 5.** The HyperW method correlates well with CLEANEX measurements for barstar. (A) $^1$H-$^{15}$N CLEANEX Fast-HSQC spectrum with $\tau_m$ = 40 ms (black) and HyperW $^1$H-$^{15}$N HMQC (red, taken from Fig. 3A) measured on $^{15}$N-labeled barstar. The post-dissolution ~355 µL sample, which contained ~1.6 mM barstar and ~1.8% $H_2O$ buffer (50 mM sodium phosphate, pH 7) was lyophilized and subsequently reconstituted in the same volume of a 90% $H_2O$. For the CLEANEX measurements indirect- and direct-domain spectral widths of 7812.5 and 2130.1 Hz were covered, using 64 $t_1$ hypercomplex increments.[34] $N_S$ = 64 scans were collected using a 131.1 ms acquisition time, and a relaxation delay of $d_1$ = 2 s. Total experimental time was ~ 5 hrs for each different mixing time $\tau_m$. For HyperW HMQC, the acquisition parameters were as in Fig. 3. All measurements were done at 50 °C on a 14.1 T Prodigy®-equipped NMR spectrometer. (B) Comparing the amide proton exchange rates $k_{UW}$ arising for different barstar residues as extracted from CLEANEX experiments[26] at 14.1 T (black squares), with the issuing HyperW HMQC sensitivity enhancements (blue circles). The sensitivity enhancements were calculated by comparing peak volumes between the HyperW HMQC spectrum (such as in Fig. 3A, red) and the thermal equilibrium spectrum measured for the corresponding sample in 90% $H_2O$ buffer. The values are averaged for two HyperW HMQC experiments, after normalizing to the $H_2O$ proton enhancement in each experiment. All measurements were done at 50 °C.

### *HyperW NMR on R17: Highlighting the unfolded state in a folded/unfolded coexisting system*

Chicken brain α-spectrin repeat 17 (R17) is a 118 residue domain, which exists in equilibrium between a well folded state (F), and an unfolded state (U).[20] The exchange dynamics between these states is very slow on the NMR timescale, with an exchange rate $k_{ex} = k_{F\rightarrow U} + k_{U\rightarrow F}$, which has an upper limit of 0.01 $s^{-1}$ at 37 °C.[20] This provides an interesting platform for assessing the "exchange filter" model put forward for barstar: as individual resonances should be observable for each of these forms, one expects that the HyperW enhancement will highlight the unfolded, exposed residues over their folded, protected counterparts. Figures 6A, 6B demonstrate that this is indeed the case, by comparing hyperpolarized and thermal data recorded at 37 °C and 2% $H_2O$ on this 13.3 kDa polypeptide, where the [U]/[F] equilibrium constant is ~1. Even a cursory investigation of the spectra shows that the HyperW procedure enhances the disordered residues appearing in the central 8-9 ppm / 118-128 ppm $^{1}H/^{15}N$ amide region, more strongly than the well-resolved peaks arising from the folded form and appearing in the periphery of this "box". The relatively good HyperW HMQC line shapes allow us to use literature data[56] in order to assign individual peaks –but only for the folded form. The majority of the unfolded peaks, unfortunately, overlap and prevent us from performing a similar complete assignment. HyperW enhancements measured for the assigned folded and the partly assigned unfolded peaks, are summarized in Figures 6C, 6D. These data confirm that the method preferentially enhances the signals from residues in the unfolded conformation than from the folded one – the average enhancement for the unfolded form is ≈100×, while for the folded one it is ≈25×. These unfolded- and folded-case values are comparable to those observed for the PhoA4 and barstar cases, respectively. For specific residues such as the indole group of W26 that can be identified in both unfolded and folded resonances, the enhancements are 35× and 10×, respectively. As enhancements are influenced by the rates of exchange and in unfolded forms these exchanges are facilitated, this is in good accord with typical amide/solvent exposure expectations.

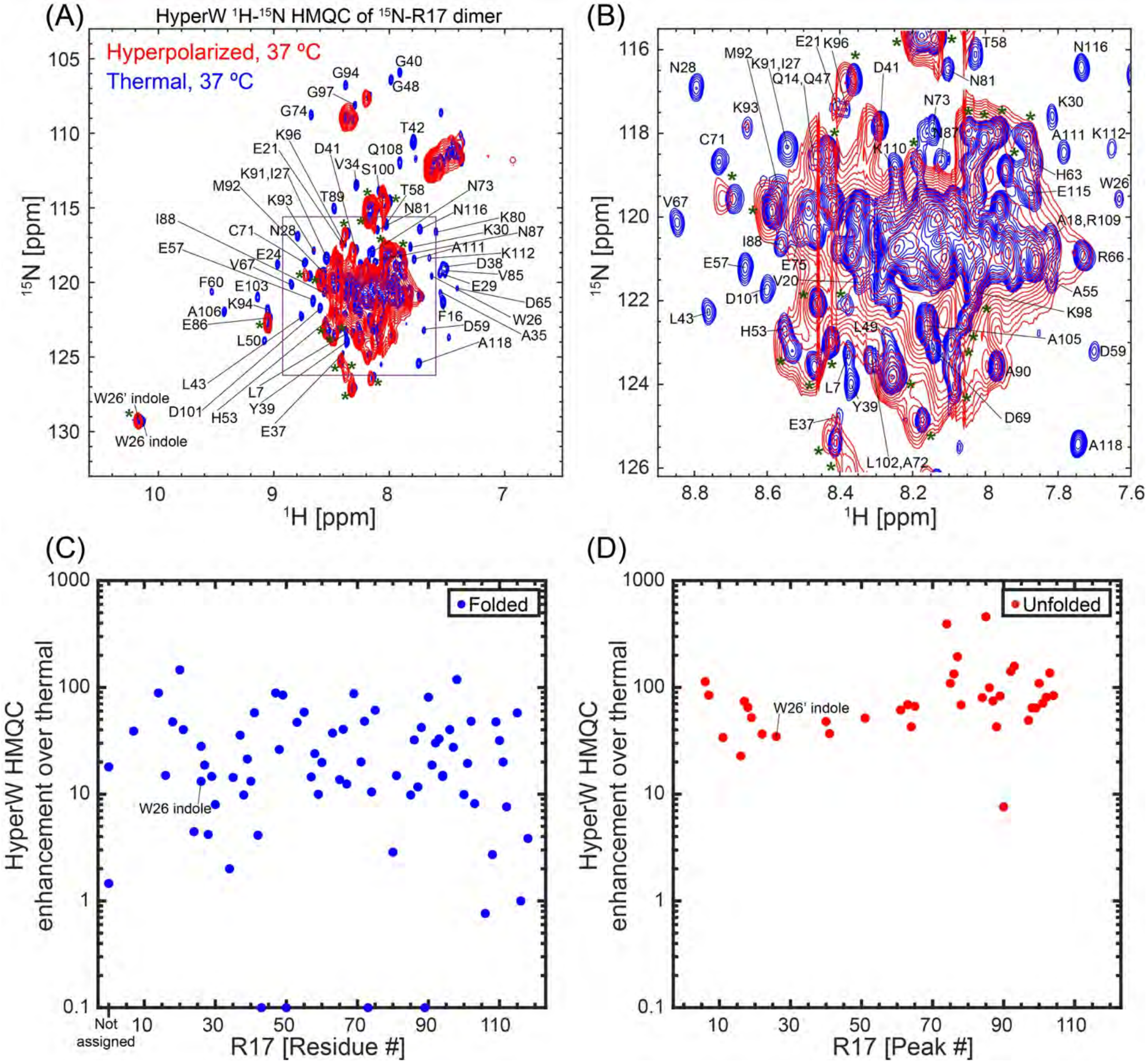

**Figure 6.** HyperW vs thermal HMQC results for R17, a protein possessing unfolded and folded conformations in slow U⇋F exchange. (A, B) Comparisons between 2D HyperW (red) and the thermal (blue) $^{1}H$-$^{15}N$ HMQC spectra measured for a $^{15}N$-R17 dimer at 37 °C. 2.8 mL of super-heated buffered $D_2O$ (50 mM HEPES, pD 7.5, 50 mM KCl) was used to dissolve an 85/15 water/glycerol pellet containing 10 mM 4-amino TEMPO. The pellet was polarized at 1.20 K for 3 hrs using microwave irradiation of 100 mW, 94.195 GHz. ~160 μL of the resulting hyperpolarized water solutions were injected into a 5 mm NMR tube containing ~140 μL of a ~1.2 mM $^{15}N$-R17 solution. Partial assignment of residues indicated by single-letter amino acid codes is done based on the BMRB entry of R17.[56] Resonances of the folded conformation are labeled with their respective assignments, resonances of the unfolded form are marked with an asterisk, and unassigned peaks are either overlapped folded and unfolded conformation residues, or residues belonging to the latter. The indole peak of W26 is assigned with a prime (') for the unfolded conformation, and without a prime for the folded one. The full spectrum is shown in (A), and a zoomed-in view (highlighted square) in (B). These spectra were recorded at 37 °C using 64 hypercomplex $t_1$ increments[34] covering indirect- and direct-domain spectral widths of 7211.5 and 1947.5 Hz; The HyperW spectrum was recorded using two phase-cycled scans per $t_1$. Total experimental time of 63 s for the hyperpolarized spectrum (acquisition time of 177.5 ms, repetition delay of 0.037 s) and 20 hrs 07 min for the thermal spectrum (256 scans with an acquisition time of 177.5 ms and repetition delay of 2 s per $t_1$ increment). (C, D) HyperW HMQC sensitivity enhancements for residues of the $^{15}N$-labeled R17 domain at 37 °C. The sensitivity enhancements were measured by comparing peak volumes between the HyperW HMQC spectrum and the thermal equilibrium spectrum measured for the same sample in (A). Sensitivity enhancements for the folded state are marked with blue circles (C), and those of the unfolded state are marked with red circles (D). Note that there is no assignment available for the unfolded state, therefore the enhancements in (D) are plotted against sequential peak numbers.

***Paradigm broken: HyperW differential enhancements of the folded and unfolded drkN SH3 domain are biased towards the former***

SH3 are small protein domains, found as modular entities in a variety of eukaryotic and viral proteins.[57-58] The SH3 domain from the *Drosophila* signal transduction protein, drkN, has an important role in behavioral neuroplasticity, in activation-dependent learning, and in memory formation.[59] It also has an interesting dynamics that was targeted by several investigations,[60-64] and showed that this 6.9 kDa polypeptide exists in equilibrium between a well folded ground state (F), and an unfolded excited state (U).[65-66] These equilibrium dynamics are slow, and thus in a simple $^{1}$H-$^{15}$N HSQC spectrum one can distinguish and assign peaks which belong to both states. Figure 7A shows a set of $^{1}$H-$^{15}$N HMQC spectra measured at different temperatures on the $^{15}$N-labeled SH3 domain from drkN, collected without hyperpolarization. These data illustrate a shift in populations in favor of the unfolded state as temperature is gradually increased; Fig. 7B highlights this with an enlargement focusing on the indole peak from Trp36 side chain, where resonances arising from F and U states at each temperature are clearly resolved, and their changing intensities can be well quantified. To further characterize this folded/unfolded equilibrium under the conditions of our study we implemented a series of ZZ-exchange NMR measurements[67] (Supporting Fig. S5), that quantify both the kinetics and thermodynamics of slow conformational exchanges such as SH3's U⇋F process. Supporting Table S1 summarizes these kinetic and population values, as derived by these measurements on SH3 at the three temperatures that we explored.

Figure 7C compares 2D HyperW vs thermal $^{1}$H-$^{15}$N HMQC spectra measured for the same, post-dissolution SH3 sample, at 50 °C and 2% $H_2O$. A mostly unfolded state dominates these spectra, whose residues (indicated by primes added to the single-letter amino acid codes) are once again significantly enhanced by the injection of hyperpolarized water. Interestingly, however, one can also observe a significant enhancement of the folded state peaks; see for instance Fig. 7D, zoomed in on the Trp36 indole peak from the folded (F) and unfolded (U) states. The reported sensitivity enhancements (Fig. 8) are calculated by comparing peak volumes between the HyperW HMQC spectrum (such as in Fig. 7C, red) and the thermal equilibrium spectrum measured for the same sample in 90% $H_2O$ buffer, after suitable rescaling to equate the proton concentrations. The degree of enhancement of these F-derived peaks is not easy to quantify from the thermal counterpart, as at an abundance of ~5.7% their visibility is limited. Furthermore, the reproducibility of hyperpolarized water injections is not perfect. Still, after $n$ = 3 injections performed under *a priori* identical conditions and after extensive signal averaging of the thermal samples, systematically higher enhancements are observed at 50 °C for this, and for many other F-state residues, than for their U-state counterparts. This is illustrated in Fig. 8 in a number of different representations, which aim at conveying the extensive experimental data that indicates that in this system, at 50 °C, water hyperpolarization enhances the majority of the assignable residues in the folded form of drkN SH3 more than in its unfolded counterpart. This anomalous behavior is observed to a much smaller extent at 37 °C, even if the folded residue enhancements there are still considerably higher than in any of the other folded proteins examined in this study.

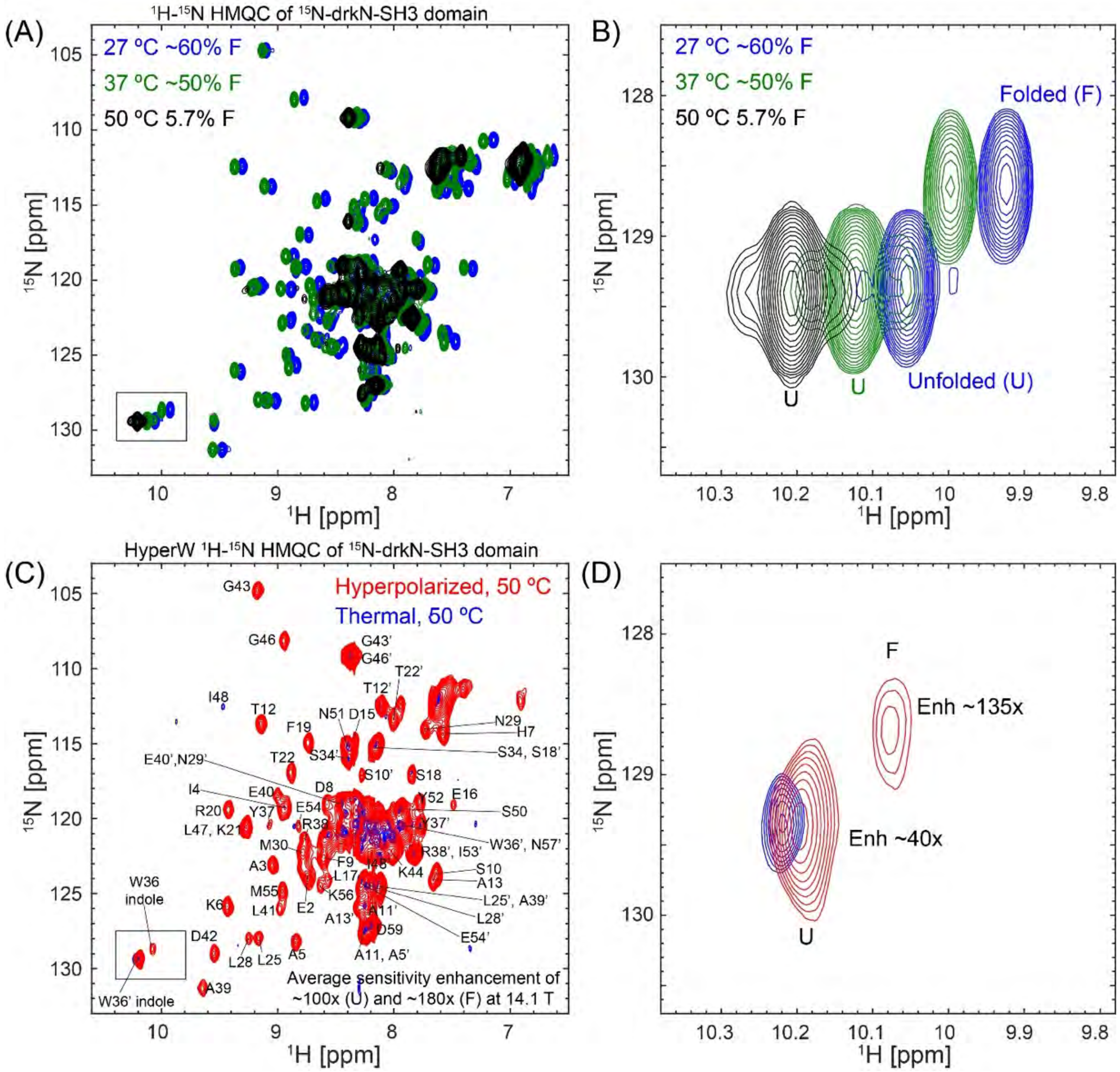

**Figure 7.** SH3 folded and unfolded states visualized by HyperW HMQC. (A) 2D $^1$H-$^{15}$N HMQC spectra measured for a ~520 μM $^{15}$N-drkN-SH3 domain in a 87.4% $H_2O$ buffer (50 mM HEPES buffer, 50 mM KCl, pH 7.5), at 27 °C (blue), 37 °C (green) and 50 °C (black). Indirect- and direct-domain spectral widths of 9014.4 and 2312.7 Hz were covered, using 64 hypercomplex $t_1$ increments.[34] The flip angle of the selective excitation was 90°, $N_S$ = 16 scans were collected using a 56.8 ms acquisition time, and a relaxation delay of $d_1$ = 2 s. Total experimental time was 1 hr 12 min. (B) Enlarged region of Trp36 indole peak (marked with a black rectangle in (A)) showing the thermally driven rise of the unfolded state. (C) Comparison between 2D HyperW (red) and thermally polarized (blue) $^1$H-$^{15}$N HMQC spectra measured for the $^{15}$N-drkN-SH3 domain. 2.8 mL of super-heated buffered $D_2O$ (50 mM HEPES, pD 7.5, 50 mM KCl) were used to dissolve an 85/15 water/glycerol pellet containing 10 mM 4-amino-TEMPO. The pellet was polarized at 1.17 K for 3 hrs 30 min using 100 mW microwave irradiation at 94.195 GHz. ~180 μL of the resulting hyperpolarized water solution were injected into a 5 mm NMR tube containing ~140 μL of a ~1.3 mM $^{15}$N-drkN-SH3 solution. Partial assignment of the various residues indicated by single-letter amino acid codes is done on the basis of Zhang et al.[65] Resonances of the folded conformation are labeled with these assignments, and resonances of the unfolded form are marked with an added prime ('). These spectra were recorded at 50 °C using 64 hypercomplex $t_1$ increments covering indirect- and direct-domain spectral widths of 7211.5 and 1947.5 Hz. The HyperW spectrum was recorded using two phase-cycled scans per $t_1$. Total experimental time of 63 s for the hyperpolarized spectrum (acquisition time of 177.5 ms, repetition delay of 0.037 s) and 13 hrs 50 min for the thermal spectrum (176 scans, acquisition time of 177.5 ms and a repetition delay of 2 s per $t_1$ increment). (D) Enlarged region of Trp36 indole peak (marked with a black rectangle in (C)).

As the results shown in Fig. 8 depart from the standard paradigm according to which unfolding should promote a more facile water/amide exchange process and hence a higher HyperW enhancement, numerous ancillary tests were performed to corroborate and further understand these findings. The simplest of them, repeated injections, gave fairly reproducible results –at least within the limits of our HyperW NMR setup, and within the resolution constraints imposed by the relatively broad unfolded spectral patterns (Supporting Table S3). CLEANEX-PM experiments were also undertaken on the post-dissolution samples, but at 50 °C they failed to provide sufficient sensitivity to measure the exchange rates of either the folded (minority) or unfolded (broadened) sites. Samples that had been analyzed by HyperW were thus lyophilized, resuspended in 90% $H_2O/D_2O$ buffer, and subjected to CLEANEX-PM analyses at 50 °C. Figure 9 summarizes representative findings of these experiments. As can be appreciated from the CLEANEX-PM spectrum measured with a mixing time comparable to the HyperW recycling delay, the buildup process only highlighted the more abundant U-derived resonances; peaks belonging to the folded state are not observed –primarily due to their low populations (these peaks were not highlighted by longer CLEANEX mixing times either). Furthermore, a relatively weak correlation (r≈0.50) was found between the $k_{UW}$ exchange rates measured in these CLEANEX-PM studies, and the corresponding sensitivity enhancements observed in the HyperW HMQC for the unfolded resonances (Fig. 9B).

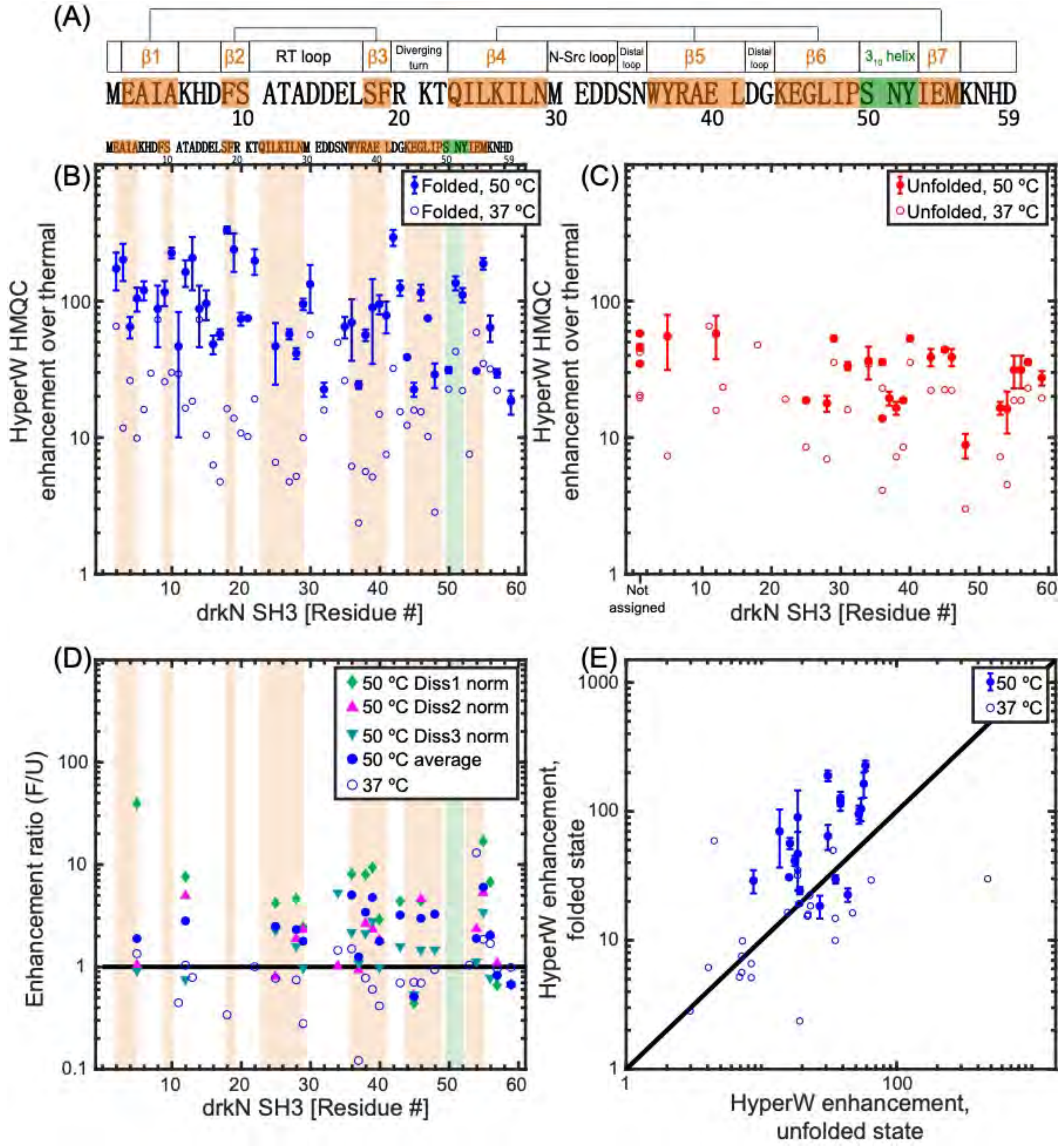

**Figure 8.** (A) 59-residue drkN-SH3 domain sequence analyzed in this study. Secondary structure elements in the folded state (as measured at 20 °C[62]) are denoted above the sequence and shaded in orange (β-strands) and green ($3_{10}$ helix). Three β-sheets are formed in this small protein, and their β-strands are connected by straight lines in the cartoon. (B, C) HyperW HMQC sensitivity enhancements for assigned residues of the $^{15}$N-labeled drkN SH3 domain at 50 °C (full symbols) and 37 °C (open symbols). The sensitivity enhancements were calculated by comparing peak volumes between the HyperW HMQC spectrum (such as in Fig. 7, red) and the thermal equilibrium spectrum measured for the same sample in ~90% $H_2O$ buffer. The values at 50 °C are averages for three nominally identical HyperW HMQC experiments after normalizing to the $H_2O$ proton enhancement in each experiment, and the "error bars" denote the spreads observed in these experiments; only residues whose identity could be verified were included in the analysis (Table S3). Sensitivity enhancements for the folded state are marked with blue circles (B), and those of the unfolded state are marked with red circles (C). (D, E) Different renderings of the observed experiments, showing the relative enhancement ratio of folded vs unfolded peaks in all the experiments recorded (D), and as correlations between the folded and unfolded enhancements observed in all the experiments at 37 and 50 °C (E). Orange and green shaded areas are drawn in (B, D) for regions which correspond to the secondary structure elements in (A).

HyperW HMQC measurements were repeated at 37 °C – where the folded state is more abundant, and the rates of U⇋F interconversion are, according to ancillary ZZ-exchange and methyl-TROSY experiments (see Fig. S5 and Tables S1 and S2 of the Supporting Information), slower. Figure 8 summarizes these results (open symbols). As can be seen both folded and unfolded peaks are now enhanced systematically less than at 50 °C; this is as expected, given the decrease in the solvent exchange rates occurring upon lowering the temperature, and the decrease in the water $T_1$ that will lead to shorter hyperpolarization lifetimes. Furthermore, individual residues are now enhanced to comparable degrees in their folded and unfolded forms. CLEANEX-PM measurements were repeated for SH3 under these conditions, to measure amide exchange rates for the resonances of the unfolded and folded forms (Supporting Fig. S6). The measured signal enhancements for the folded state of drkN-SH3 at 37 °C correlate well (r=0.85) with solvent-amide exchange rates measured by CLEANEX-PM at this temperature (Fig. S6A), while for the unfolded state this correlation is weaker (r=0.49, see Fig. S6B). The enhancements observed at 37 °C are in agreement with the expectations deriving from the HyperW examples discussed above, as the enhancements for folded state residues at this temperature are equal to or smaller than for their unfolded counterparts. This lifts the need for an explanation of anomalous folded-vs-unfolded enhancements at this lower temperature, but does not shed light on the behavior observed at 50 °C.

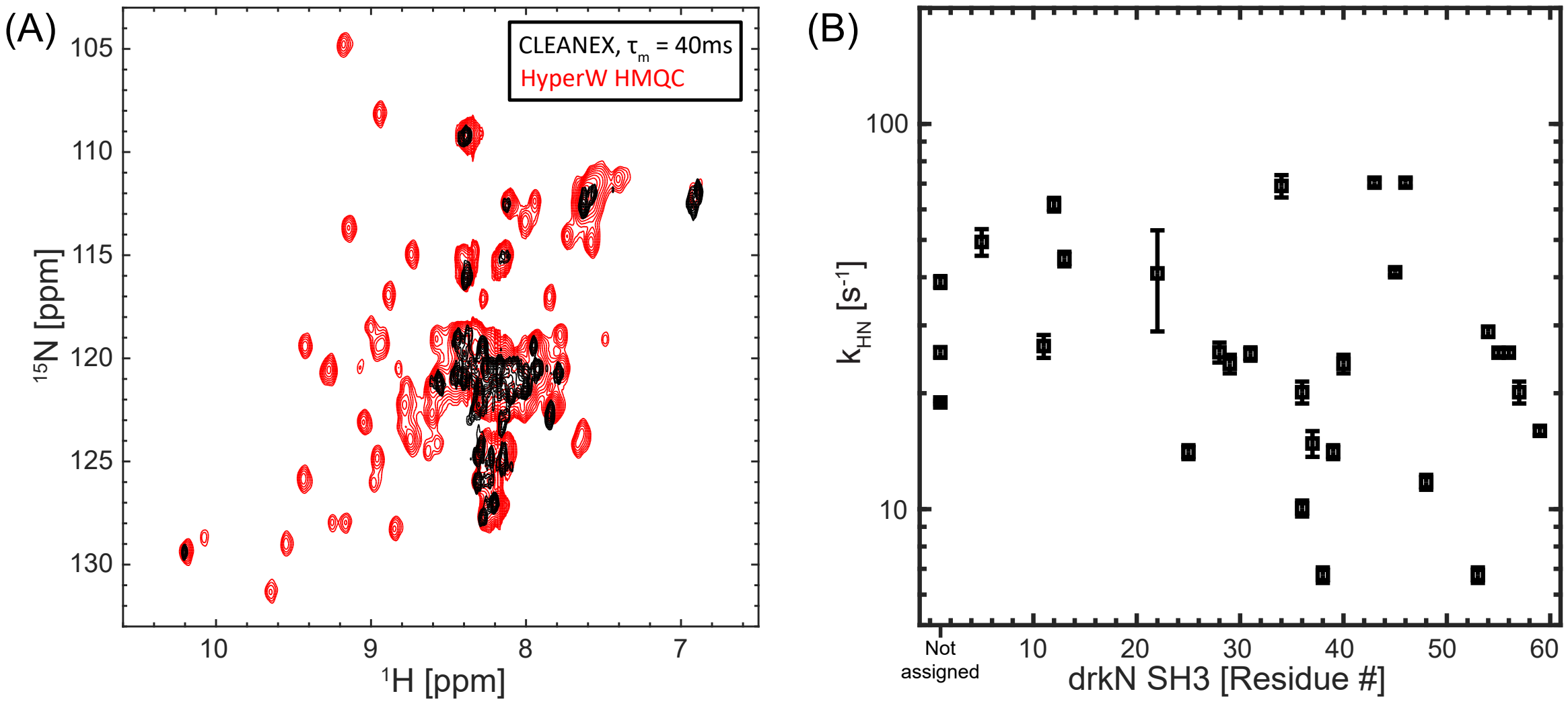

**Figure 9.** (A) $^1$H-$^{15}$N CLEANEX fast-HSQC spectrum with $\tau_m$ = 40 ms (black) and HyperW $^1$H-$^{15}$N HMQC (red, taken from Fig. 7C) measured on $^{15}$N-labeled drkN SH3. Notice how HyperW enhancements appear to correlate with CLEANEX-PM measurements for the unfolded state of the drkN SH3 domain. For the CLEANEX-PM measurements (black), one of the post-dissolution samples containing 160 µM drkN SH3 and ~2.4% $H_2O$ buffer (50 mM HEPES, 50 mM KCl, pH 7.5) was lyophilized and subsequently reconstituted in the same volume with 90% $H_2O$. For CLEANEX the indirect- and direct-domain spectral widths were 7211.5 and 2069.2 Hz, covered using 64 $t_1$ hypercomplex increments and STATES acquisition.[34] $N_S$ = 128 scans were collected using a 142.0 ms acquisition time, and a relaxation delay of $d_1$ = 2 s. Total experimental time was ~ 10 hrs for each different mixing time $\tau_m$. For the HyperW HMQC, acquisition parameters were as in Fig. 7C. All measurements were done at 50 °C on a 14.1 T Prodigy®-equipped NMR spectrometer. (B) Amide proton exchange rates $k_{UW}$ arising for different drkN SH3 residues in the unfolded state as extracted from CLEANEX-PM experiments at 14.1 T, 50 °C (black squares).

A feature that distinguishes SH3's 50 °C behavior from both its behavior at 37 °C and the R17 case, concerns the presence of a relatively fast U⇋F interconversion between a dominant U and a minority F state. At these conditions the folded form corresponds to what is normally considered to be an "invisible" state,[68] which is only made visible here by the unusually large enhancements brought to its amide peaks by the hyperpolarized water injection. This suggests the possibility of an alternate route to the water(H) ⇋ amide(H) exchange facilitating the HyperW HMQC enhancement, along the lines shown in Scheme 1. In this case the folded form is hyperpolarized by two concurrent processes: one where the water protons undergo direct chemical exchange with the amides of the folded state, and another where this exchange happens with the protons of the highly populated unfolded state –and then this unfolded state undergoes a conformational conversion into the lowly populated folded form.

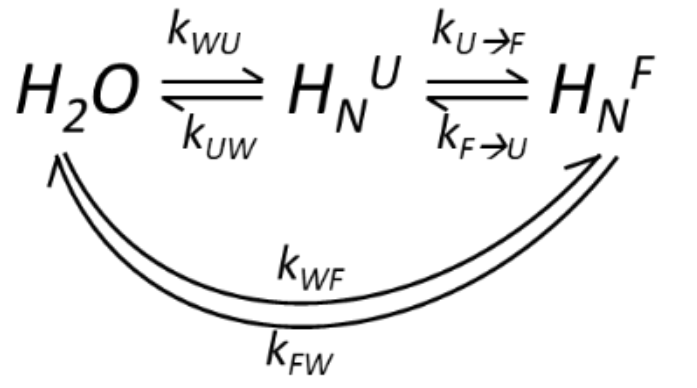

**Scheme 1.** Potential exchange processes defining HyperW experiments on drkN SH3 domain. U and F denote a residue's unfolded and folded conformations; $k_{WU}$, $k_{WU}$ are the exchange rates of the water protons with the amides in the unfolded and folded states, and $k_{XW}$ (X = U,F) are the rates of the backward reactions. $k_{U\to F}$ and $k_{F\to U}$ are the rates of the U⇋F protein interconversion.

A theoretical Bloch-McConnell exchange model was developed to test whether these additional dynamics could explain the anomalous enhancement of the folded over the unfolded residues; calculations showed that the enhancements measured for the folded state residues could then indeed be larger than for the unfolded state– but only if the solvent exchange rates for these folded residues are faster than for their unfolded counterparts (Figs. 10A,10B). In search for an alternative that would demand less radical assumptions, the exchange model was expanded to include potential effects of different cross-relaxations among the folded and unfolded states. Accounting for this required computing water, amide and aliphatic magnetizations $<H_2O>_z$, $<H_N^U>_z$ and $<H_N^F>_z$, $<H_C^U>_z$ and $<H_C^F>_z$, as solutions of a system of Bloch-McConnell-Solomon equations. For the sake of completeness, we included into this model the possibility that the 33 labile sidechain sites in this 59 residues peptide –representing hydroxyls, guanidinium and amines– might also be enhanced by exchanges with the hyperpolarized water, and transfer their hyperpolarization via cross relaxation to the targeted amide sites. While a more complete account of this model is given in the Supporting Information, the overall system of equations that we considered was[28]

$$\frac{d}{dt}\begin{pmatrix} \langle H_N{}^F\rangle_z(t)-\langle H_N{}^F\rangle_z(eq) \\ \langle H_N{}^U\rangle_z(t)-\langle H_N{}^U\rangle_z(eq) \\ \langle H_C{}^F\rangle_z(t)-\langle H_C{}^F\rangle_z(eq) \\ \langle H_C{}^U\rangle_z(t)-\langle H_C{}^U\rangle_z(eq) \\ \langle H_X{}^F\rangle_z(t)-\langle H_X{}^F\rangle_z(eq) \\ \langle H_X{}^U\rangle_z(t)-\langle H_X{}^U\rangle_z(eq) \\ \langle H_2O\rangle_z(t)-\langle H_2O\rangle_z(eq) \end{pmatrix} =$$

$$\begin{pmatrix} -r_F & k_{UF} & \sigma_F & 0 & \sigma_{XF} & 0 & k_{WF}+\sigma_{WF} \\ k_{FU} & -r_U & 0 & \sigma_U & o & \sigma_{XU} & k_{WU}+\sigma_{WU} \\ \sigma_F & 0 & -R_1^{H_C F} & k_{UF} & \sigma_{XF} & 0 & \sigma_{WF} \\ 0 & \sigma_U & k_{FU} & -R_1^{H_C U} & 0 & 0 & \sigma_{WU} \\ \sigma_{XF} & 0 & \sigma_{XF} & 0 & -R_1^{H_X F} & k_{UF} & k_{WX}+\sigma_{WFX} \\ 0 & \sigma_{XU} & 0 & \sigma_{XU} & k_{FU} & -R_1^{H_X U} & k_{WX}+\sigma_{WUX} \\ k_{FW}+\sigma_{FW} & k_{UW}+\sigma_{UW} & 0 & 0 & k_{XW}+\sigma_{WFX} & k_{XW}+\sigma_{WUX} & -r_W \end{pmatrix} \begin{pmatrix} \langle H_N{}^F\rangle_z(t)-\langle H_N{}^F\rangle_z(eq) \\ \langle H_N{}^U\rangle_z(t)-\langle H_N{}^U\rangle_z(eq) \\ \langle H_C{}^F\rangle_z(t)-\langle H_C{}^F\rangle_z(eq) \\ \langle H_C{}^U\rangle_z(t)-\langle H_C{}^U\rangle_z(eq) \\ \langle H_X{}^F\rangle_z(t)-\langle H_X{}^F\rangle_z(eq) \\ \langle H_X{}^U\rangle_z(t)-\langle H_X{}^U\rangle_z(eq) \\ \langle H_2O\rangle_z(t)-\langle H_2O\rangle_z(eq) \end{pmatrix} \quad (2)$$

Diagonal elements in the matrix above are given by $r_F = k_{FW} + k_{FU} + R_1^{H_N F} + \frac{1}{T_1^F}$, $r_U = k_{UW} + k_{UF} + R_1^{H_N U} + \frac{1}{T_1^U}$, $R_1^{H_C F}$, $R_1^{H_C U}$, $R_1^{H_X F}$, $R_1^{H_X U}$ and $r_W = \frac{1}{T_1^W} + k_{WU} + k_{WF} + 2k_{WX}$. In these expressions the rates $k_{FU}$ and $k_{UF}$ represent the exchange rates between the folded and unfolded states (Scheme 1); $R_1^{H_N F}$, $R_1^{H_N U}$, $R_1^{H_X F}$, $R_1^{H_X U}$, $R_1^{H_C F}$ and $R_1^{H_C U}$ are the auto-relaxation rates of the amide, the labile sidechain and the aliphatic protons in folded and unfolded states (including dipolar interactions between the $^{15}N$ and $H_N$, $H_N$ and two $H_C$s, $H_N$ and a suitably weighted $H_X$, $H_N$ and the hyperpolarized water, as well as between pairs $H_C$-$H_C$ of aliphatic protons). Notice that an additional rate $1/T_1$ was added to the relaxation terms of each amide proton and of the water, to account for other potential effects –arising, for instance, from the residual radical. The various σ's in Eq. (2) represent in turn the cross-relaxation rates among the various proton populations. $\langle H_2O\rangle_z(eq)$, $\langle H_N\rangle_z(eq)$, $\langle H_X\rangle_z(eq)$ and $\langle H_C\rangle_z(eq)$ are the water, the amide (*N*), the labile sidechain (*X*) and the aliphatic (*C*) proton magnetizations at thermal equilibrium. For both hyperpolarized and thermal calculations the equilibrium polarizations were scaled according to *a priori* known molar fractions:

$$\langle H_2O\rangle_z(eq) = \frac{x_{H_2O}}{x_{H_N{}^F}} \equiv X_{WF}\ ;\ \langle H_N{}^U\rangle_z(eq) = \frac{x_{H_N{}^U}}{x_{H_N{}^F}} \equiv X_{UF};\ \langle H_N{}^F\rangle_z(eq) = 1; \quad (3)$$

the population of the exchangeable sidechain protons was reduced, to match the ratio between these nuclei and the amides in the protein. Exchange rates are also related to each other by these water and protein molar fractions: $k_{UW} = \frac{X_{H_2O}}{X_{H_N}{}^U} \cdot k_{WU}$; $k_{FW} = \frac{X_{H_2O}}{X_{H_N}{}^F} \cdot k_{WF}$. Water relaxation times were estimated from independent experiments, while kinetic parameters for the U⇋F interconversion process were extracted from the ZZ-exchange experiment shown in the Supporting Information and recorded at 50 °C (Supporting Fig. S5 and Table S1). With all this information, and using additional known parameters and standard assumptions (delay between scans, number of $t_1$ increments, number of signals averaged scans, coherence transfer efficiencies, etc.; see Supporting Information for a full derivation of this model and the assumptions involved), the relative enhancement of the HyperW vs the thermal HMQC experiments were cast in terms of three variables: The initial enhancement factor $\varepsilon = \frac{\langle H_2O\rangle_Z(0,hyp)}{\langle H_2O\rangle_Z(thermal)}$ of the injected hyperpolarized water over its thermal magnetization –a parameter that affected the enhancement of all residues, in both the folded and unfolded states, homogeneously; the $k_{UW}$ rate of exchange between water and an unfolded residue; and the rate of exchange between water and a folded residue ($k_{FW}$). Numerical calculations based on Eq. (2) were carried out for hyperpolarized and for thermally polarized HMQC acquisitions for sets of exchange rates $k_{UW}$, and $k_{FW}$, and the ensuing signal enhancement (*Enh*) was determined for each pair of residues in the set as $Enh_U = \frac{S_{U,Hyp}(k_{UW},k_{FW})}{S_{U,Thermal}(k_{UW},k_{FW})}$ ; $Enh_F = \frac{S_{F,Hyp}(k_{UW},k_{FW})}{S_{F,Thermal}(k_{UW},k_{FW})}$.

Figure 10 shows a summary of these calculations, which focuses on illustrating how the solvent exchange rates $k_{UW}$, $k_{FW}$ will affect the per-scan enhancement of different sites in drkN SH3's unfolded and folded conformations. For a range of intrinsic relaxation times $T_1^F$, $T_1^U$ and for typical water enhancement factors ($\varepsilon \approx 500$), these plots show two surfaces that intersect when $k_{UW} \approx k_{FW}$ – with some parameters being fixed as per the SH3 experiments and the ancillary independently measured data, and others varied so as to illustrate their effects. It follows from this model that the enhancements measured on the folded state residues could indeed be very large –even larger than for the unfolded state– but in the absence of water-derived cross-relaxation effects, this would require that the solvent exchange rates for these folded residues be faster than for their unfolded counterparts. Indeed, the U⇋F interconversion, intra-residue cross-relaxation effects and *ad hoc* $1/T_1$ rates will affect the symmetry of the folded and unfolded state enhancements slightly, but for the values measured independently for $k_{UF}$, $k_{FU}$, this asymmetry is relatively small (Figs. 10A, 10B). Only if equilibrium signals recorded with fixed repetition times $TR_{Hyp}$ = 0.24 s and $TR_{TE}$ = 1.21 s. Numerically simulated per-scan enhancements for the unfolded and folded conformations are plotted as 3D surfaces and as a function of exchange rates $k_{UW}$ and $k_{FW}$, for relevant ≤60 $s^{-1}$ values. (A,B) Results expected for different intrinsic relaxation times of the folded ($T_1^F$) and unfolded ($T_1^U$) states (indicated on the top of each panel), assuming that cross-relaxation processes occur solely within the $H_N$, $N_H$, and two aliphatic sidechain protons $H_C^1$ and $H_C^2$ system. Note that when a larger intrinsic relaxation rate $1/T_1^F$ is assumed, the per-scan enhancements for the folded state will be larger for slightly slower $k_{FW}$, but these effects are small. (C,D) Effects introduced when the possibility of cross relaxation from the hyperpolarized water is added to the model in (A), assuming the indicated correlation times $\tau_c$ of the folded and unfolded states. The only significant bias of the HyperW enhancements provided by the exchange processes (A,B), arises when assuming suitably different correlation times and short inter-atomic water-amide distances (panel D). (E,F) Same as models (C,D), but now incorporating the possibility of having cross-relaxation between the amides and a labile $^1$H (X), which could be part of an hydroxyl, amino or guanidinio sidechain. For simplicity the $\tau_c$'s used to model these additional relaxation processes were assumed as for the structural waters, and the rates of exchange with the solvent were assumed 30 Hz for all forms. Again, notice that very short internuclear distances would be required for these processes to have a noticeable effect.

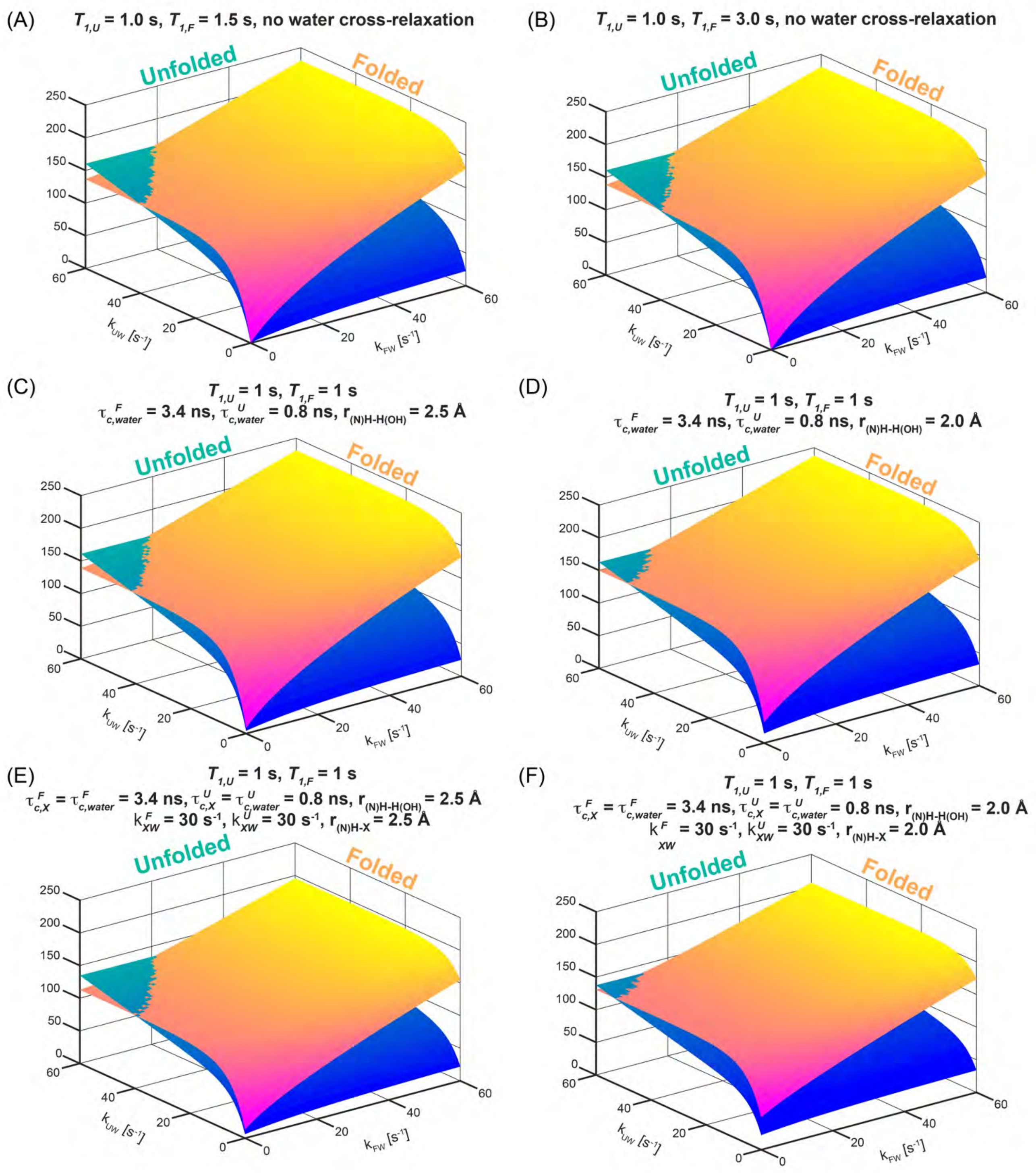

**Figure 10.** Relative HyperW/thermal enhancement per scan predicted by the numerical solutions of Eqs. (S6)-(S12), for a protein residue subject to the 2D $^1$H-$^{15}$N HMQC sequence depicted in Fig. S2 (Supporting Information). Calculations were repeated for thermal ($\varepsilon = 1$) and hyperpolarized ($\varepsilon = 500$) water scenarios as a function of exchange rates $k_{UW}$ and $k_{FW}$. Additional assumptions included $T_1^W = 15$ s, [$H_2O$] = 0.92 M (to account for a dilution to 1.7% after dissolution), [protein] = 0.59 mM, $p_U = 94.3\%$ in the hyperpolarized experiment and 96% in the thermal (to account for equilibrium differences in protonated and deuterated solvents), $p_F = 5.7\%$ in the hyperpolarized experiment and 4% in the thermal one (Table S2), $k_{UF} = 1.9$ s$^{-1}$, $k_{FU} = 31.4$ s$^{-1}$. The correlation times $\tau_c$ for the folded and unfolded states were assumed to be equal to 3.4 ns and 0.8 ns, respectively. The number of scans per increment were 2 and 128 for the hyperpolarized and thermal experiments and $N_1 = 128$ increments for both cases. Other considerations regarding the auto-relaxation and cross-relaxation are as detailed in the Supporting Information. Enhancements were calculated by taking the ratio of the expected HyperW and thermal

the $T_1^F$ value is for some reason much larger than $T_1^U$, will a slight bias towards the folded-form enhancement arise (as a result of the partial saturation of the thermal signal used as reference, a condition that was not met in our experiments) and hence the *apparent* enhancement of the folded site will look larger than of its unfolded counterpart. When strong water-associated cross-relaxation effects deriving from either the hyperpolarized water itself or from labile sidechain protons that have been hyperpolarized by the water are included, however, (e.g., Figs. 10D, 10F), the experimental data can also be reproduced if it is assumed that $k_{FW} < k_{UW}$. Notice, however, that even under these assumptions –which bias cross-relaxation enhancements towards the folded form by virtue of relatively long correlation times and short internuclear distances between the hyperpolarization sources and the targeted protons– the maximal F/U ratios reached under the $k_{FW} < k_{UW}$ condition amount to 10s %.

## Discussion and Conclusions

The injection of hyperpolarized water in precise aliquots into a regular NMR setting followed by the acquisition of high-resolution 2D data was applied to a wide range of protein structures, and shown to be a technique that can serve two main purposes. On one hand it can help to sensitize 2D HMQC NMR experiments, to the point of highlighting lowly-populated "invisible" states that would be hard to observe in equilibrium with their more populated states.[68-69] On the other hand, the experiment affords enhancements that can in general be translated into insight about relative solvent exchange for different residues within the same sample/protein. This could be important, as given reasonably well-known parameters including the hyperpolarized water enhancement and the effective $T_1$ relaxation decays, absolute values of water/amide exchange rates could also be derived. These features were explored here using an array of representative protein systems, chosen to illustrate a variety of scenarios. The largest enhancements were observed, as could have been expected, for the case of disordered proteins like the PhoA4 fragment, for which nearly all residues exhibited enhancements ≥100× –and several residues exceeded 500-fold enhancement values. Also in agreement with the aforementioned exchange-dominated model was the behavior of barstar, a well-folded protein that exhibited correspondingly smaller enhancements. Notable heterogeneities in the enhancement of the different barstar residues were noted, yet these correlated well with their readiness for water exchanges as evidenced by CLEANEX-PM measurements. Previous reports suggested that even though barstar holds a well-defined three dimensional structure, it is still dynamic and flexible;[37, 45, 51,70-71] this could help to rationalize the observed HyperW/CLEANEX NMR behavior in terms of local disorder. In fact previous H/D exchange studies investigate amide/water exchange rates for different barstar residues, and found that these exchange rates correlate with calculated relative surface accessibility (RSA).[71] It is generally accepted that in folded proteins, protons residing in flexible loops will be the most surface-exposed, while protons in secondary structure elements will be involved in hydrogen bonding or buried in the protein core, and hence their exchange rates would be slower.[72] While the HyperW enhancements observed for barstar are higher for loop regions and exposed amides (Fig. 4), several residues do not follow this correlation: most of these belong to an α-helix and are apparently involved in hydrogen bonds, yet still exhibit high enhancements. This could, however, still be explained in terms of the solvent accessibility of these residues, as they might reside in a more surface-exposed side of the α-helix.

Attention was then turned to two proteins featuring coexisting folded and unfolded states. One of these, R17, behaved within expectations: The folded peaks of R17 showed enhancements in the 1-100× range, while the same residues in the unfolded form showed enhancements in the 10-500-fold range (Fig. 6). While the resolution of this U form in the HMQC experiment wasn't sufficiently high to permit residue-specific analyses, the trends respected the behavior described above for amide exchange in unfolded and folded proteins; this was as expected, given the relatively slow interconversion between R17's F and U forms. By contrast the second system analyzed, drkN SH3, revealed an anomaly: for the majority of the assignable residues, consistently larger HyperW enhancements were observed in the folded than in the unfolded states at 50 °C. Just as the residues' enhancements were heterogeneous in each of the previously discussed systems, a distribution again characterized the individual residues' enhancements in both U and F states –in SH3's anomaly arises from the fact that the HyperW enhancements at 50˚C were in general larger for the folded state *of the same amino acid,* than for its unfolded counterpart. These anomalous trends consistently emerged when examining both the post-dissolution samples, as well as lyophilized post-dissolution samples that had been reconstituted in per-protio solvents for the sake of improving the sensitivity, even after their populations had been suitably corrected to account for solvent differences. CLEANEX measurements shed little light on the origin of this behavior: for the 50 °C case, [F] << [U]; this, plus CLEANEX's limited sensitivity, prevented the characterization of the minority, "invisible" F-state behavior (while, however, still allowing measurements of the unfolded state's behavior; see Fig. 9).

In an effort to explain how hyperpolarized water could enhance certain residues more in their folded than in their unfolded states, a theoretical model based on Bloch-McConnell's and Solomon's equations was developed. This relied on independently-measured relaxation times, on U⇋F kinetic and thermodynamic equilibrium parameters that were also independently measured, and on a variety of potentially concurrent tumbling-induced self- and cross-relaxation phenomena. The correlation times of folded and unfolded proteins were estimated based on values for chains of similar size; the main unknowns in this model were thus the rates of folded- and unfolded-state exchanges with water, and the extent of water-protein and intra-protein cross relaxations. With this model we explored whether an amide proton in the U-form could gain magnetization from the hyperpolarized water, but then "lose it" rapidly to a minority F-state that would then display unusually large enhancements as a result of combining multiple sources of hyperpolarization. These effects, however, were not significant. Our model then considered whether cross-relaxation of the amides to other, non-exchangeable (and therefore not hyperpolarized) protons in the protein, could bias these measurements and result in an artificially higher F-form enhancement. These effects, however, ended up leading to bigger losses for the more structured folded form than for the more mobile unfolded form: if there is any bias derived from these effects, it should thus be working against the apparent enhancements observed for the F residues. Inclusion of ancillary *ad hoc* $T_1$ terms did not have much influence either. The model was therefore expanded to allow for drkN SH3 amide proton enhancements to arise from other sources, including the possibility of having differential folded/unfolded cross-relaxation between the amide groups and both the hyperpolarized solvent, as well as between the amide groups and labile sidechain protons. The former, in particular, might lead to sizable contributions if structural-like hydration waters are involved.[76,77] When assuming that correlation times were sufficiently short for the unfolded and long for the folded forms, and that the intermolecular $^1H$-$^1H$ distances were sufficiently short to ensure a strong Overhauser interaction, these additions predicted that HyperW enhancements could indeed be larger for the folded than for the unfolded forms –while still respecting the $k_{FW} \leq k_{UW}$ condition (Figs. 10D-10F). The resulting enhancement differences, however, were still relatively small: ≤50% for the best $k_{FW} = k_{UW}$ case, compared to the differential enhancement factors of ca. 200-400% that are observed for numerous residues at 50 °C (Table S3).

In view of this, other potentially confounding factors were explored. One of them concerned the possibility of a thermally induced drkN SH3 degradation and/or aggregation, which were found to occur at 50 °C but only over 48 hs; these, however, are not relevant timescales for the ca. minute long times involved in our NMR measurements. Another potentially important factor that was considered, concerned potential miscalibrations of the temperatures assumed in the HyperW experiment: as lower HyperW measurement temperatures would mean larger-than-assumed folded/unfolded drkN SH3 ratios in the sample this could lead, after normalizing by intensities measured on a correctly set, thermally polarized 50 °C sample, to a bias in the ensuing folded/unfolded enhancement calculations. While no such artifacts were observed in calibration measurements (data not shown), we also relied on SH3's own high temperature dependence, to evaluate what the effects of dealing with lower-than-expected post-mixing temperature would be: comparisons against variable-temperature drkN SH3 HMQC data showed that, post-injection, HyperW sample temperatures reached the targeted 49-50 °C temperature within ca. 10 sec (Supporting Fig. S1). The various HyperW data sets collected in this study were still reevaluated under the possibility that the sudden injection process dropped the sample's temperature to 47 °C but, as shown by supporting Figure S7, this would still leave, within experimental errors– the majority of assignable folded peaks in the HyperW spectra equally or more enhanced than their unfolded counterparts. As mentioned earlier, also the population imbalances that may arise upon comparing folded/unfolded equilibria in mostly deuterated (e.g., HyperW) and mostly protonated (thermal) water were considered; these were also measured via ancillary ZZ-exchange and methyl-TROSY experiments (Supporting Information, Tables S1 and S2) and their effects included in all enhancement estimations.

When examining which folded-form drkN SH3 sites showed the largest HyperW enhancements (Fig. 8), residues at or near disordered loops stood out: for these cases nearly 300× enhancements could be measured, vis-à-vis ≈100-fold enhancements for their unfolded counterparts (see Supporting Information Table S3 for a summary of drkN SH3's 50 °C results). This might explain why these residues are enhanced more than other amides in better folded regions –or in other folded systems we have examined. It still leaves the question, however, of how *the same* residue can be more readily enhanced by hyperpolarized water in a folded than in an unfolded form. Although the solvent/amide and labile-sidechain/amide cross-relaxation arguments made above could partly explain this behavior, it is hard to discard completely the role that amide-solvent exchange rates could play in this anomaly. Solvent-protein exchange measurements have been the focus of decades of systematic studies,[73-75] with NMR- and mass-spectrometry-based H/D exchange measurements being the most established forms for measuring them.[76-81] These solvent/amide exchange measurement, which are clearly related to the HyperW NMR measurement, have in turn been intimately linked with the degree of folding (or intermolecular binding) of a protein.[77, 82] This derives from the reasonable assumption that the more easily that water can access a specific amide moiety, the faster the rate of exchange with water will be.[83-84] A change in the rate of solvent exchange will thus reflect a change that the residue in question experiences vis-à-vis solvent accessibility —up to a maximum rate given by the exchange of the isolated amide (for instance, in a model dipeptide structure). Decades of H/D exchange studies have also revealed that many factors beyond solvent accessibility may influence a particular amide's solvent exchange rate, and change it by factors of up to a billion-fold. Foremost among these factors are the group's local acidity,[85-90] the effective electrostatic charge of the residue involved,[87-89,91-96] and the electrostatic shielding imposed by a residue's neighbors.[89, 97-98] On the basis of these very strong influences, it has been hypothesized, and even predicted by numerical methods,[89,99] that anomalous cases may arise where rates of H/D exchange do not correlate with exposure to the solvent –and hence with a residue's degree of folding. To the best of our knowledge, however, such predictions have not heretofore been experimentally detected. In this respect, the HyperW method provides a unique experimental window that could enable the discovery of such instances: by its very nature it probes the solvent accessibility directly and in very short timescales; it does so in a residue-by-residue fashion; it provides the ability to discriminate between peaks arising from coexisting folded and unfolded forms; and by virtue of its enhanced sensitivity it enables one to see minority states that under normal conditions would be invisible. As such it allowed us to monitor enhancements of SH3's folded and unfolded states under conditions that are at the threshold of total unfolding. It remains to be seen whether additional experiments can be devised, that shed further light on the origins of the unusually high HyperW enhancements displayed by the folded residues over their unfolded counterparts.

The present study presented some of the promising avenues opened by HyperW NMR in protein research. It verified that even in its present form it can be used to sensitize the spectra of IDPs by several hundred-fold. It showed that even proteins like barstar, which are typically considered to be essentially folded, can also experience substantial enhancements that inform about the local structure and dynamics of the protein. Most intriguingly, it also provided a new experimental tool to examine coexisting folded and unfolded protein states –even when one of these is present at what are normally "invisible" concentrations. Still, numerous additions could extend further the analytical power of this approach to solution-state protein NMR spectroscopy. Aspects in need of improvements from the DNP standpoint include increasing the volume and the hyperpolarization of the water,[100] eliminating the polarizing radical[101-102] and –foremost of all– reducing the dilution experienced by the hyperpolarized water. Additional improvements investigated partially in this work, like the reliance on non-uniform sampling schemes, could also facilitate higher sensitivity, higher resolution,[48-49] and extensions to higher dimensionalities.[103-105] Several of these advances are currently in the making, in the hope of revisiting the behaviour of multiple protein unfolded/folded equilibria and of probing solvent accessibility in more complex interacting systems.

**Acknowledgment:** This work would not have been possible without the technical assistance of the late Koby Zibzener. We are also grateful to Dr. S.F. Cousin for assistance in the cross-relaxation calculations, to Dr. Shira Albeck (ISPC, Weizmann Institute of Science) for the barstar protein, and to Ms. Ivana Petrovic for expression of $^{13}C$-labeled drkN-SH3. This work was supported by the Kimmel Institute for Magnetic Resonance (Weizmann Institute), the EU Horizon 2020 program (Marie Sklodowska-Curie Grant 642773 and FET-OPEN Grant 828946, PATHOS), Israel Science Foundation Grant 965/18, and the Perlman Family Foundation. RR was also supported by the Israel Science Foundation Grant 1889/18 and a Minerva Foundation Research Grant.

## ASSOCIATED CONTENT

**Supporting Information.** Additional details are given about the experimental procedures and pulse sequences used; controls on the post-injection temperatures; additional measurements on PhoA4; additional ZZ-exchange and methyl-TROSY measurements on SH3 at 37 and 50 °C; and details on the model accounting for three-site exchanges with cross-relaxation. The Supporting Information is available free of charge on the ACS Publications website (PDF).

## AUTHOR INFORMATION SECTION

[&]Current address: Institute of Biological Chemistry, Faculty of Chemistry, University of Vienna, Währinger Straße 38, 1090 Vienna, Austria.

[$] Current address: Department of Biochemistry, Duke University School of Medicine, 307 Research Drive, Durham NC 27710, USA

* Corresponding author. E-mail: lucio.frydman@weizmann.ac.il

Graphic Table of Contents

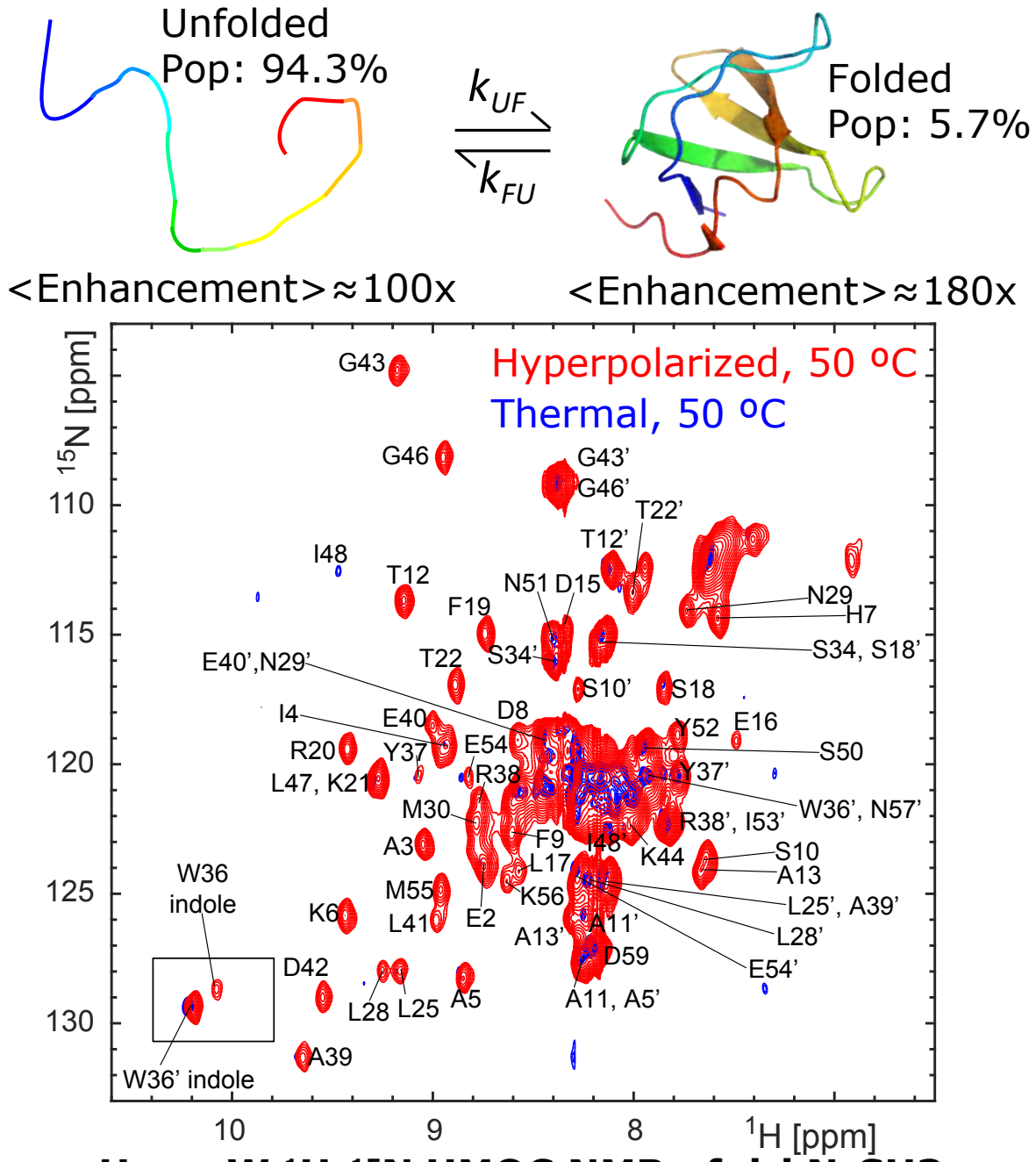

**HyperW $^{1}H$-$^{15}N$ HMQC NMR of drkN-SH3**

Supporting Information for

# Assessing site-specific enhancements imparted by hyperpolarized water in folded and unfolded proteins by 2D HMQC NMR

Or Szekely[†,$], Gregory Lars Olsen[†,&], Mihajlo Novakovic[†], Rina Rosenzweig[‡], Lucio Frydman*[,†]
Departments of [†]Chemical and Biological Physics and [‡]Structural Biology, The Weizmann Institute of Science, 234 Herzl Street, Rehovot 760001, Israel
[&]Current address: Institute of Biological Chemistry, Faculty of Chemistry, University of Vienna, Währinger Straße 38, 1090 Vienna, Austria.
[$] Current address: Department of Biochemistry, Duke University School of Medicine, 307 Research Drive, Durham NC 27710, USA

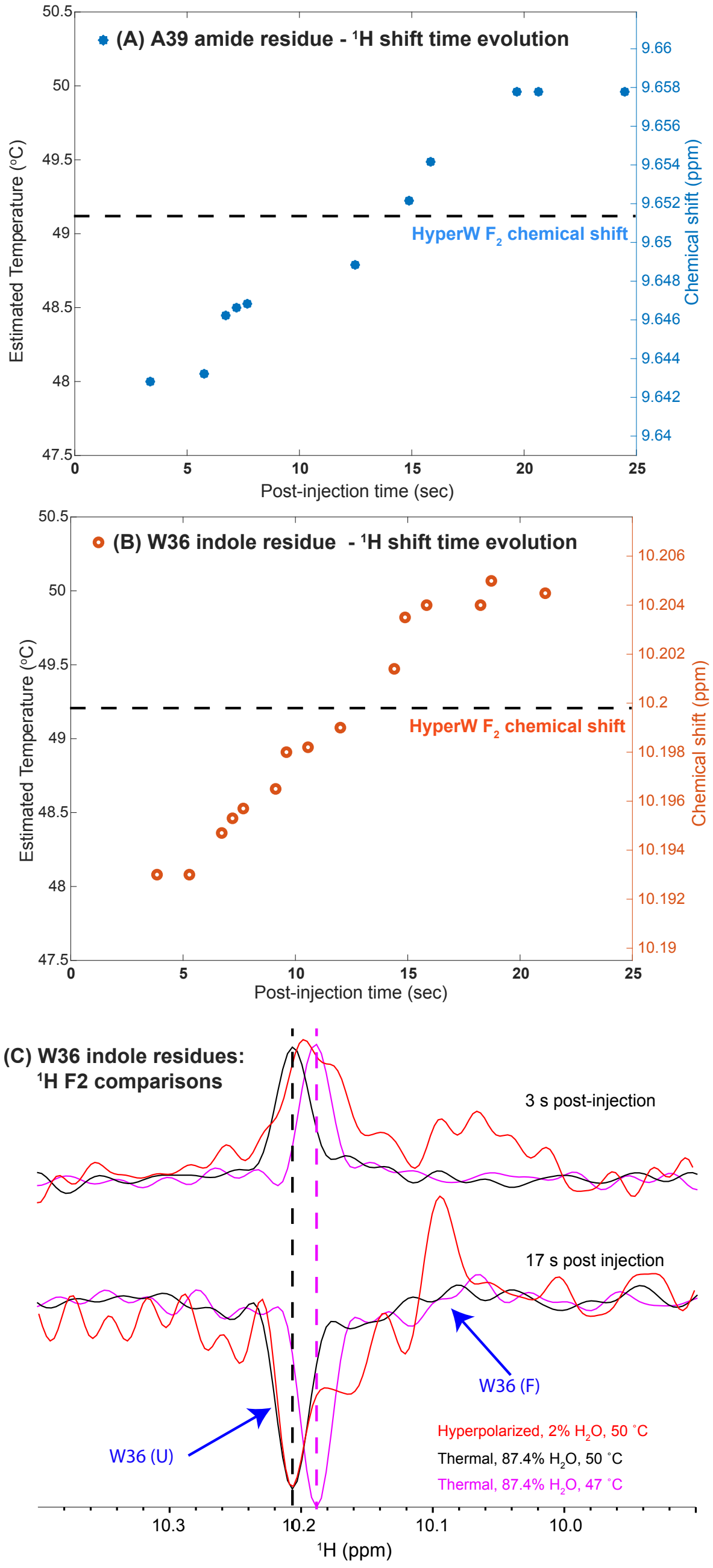

**NMR Spectroscopy.** Sample temperature is an important aspect for the claims made in this study. In principle temperatures should be quite stable and uniform throughout the injections: both the protein solution waiting inside the NMR and the hyperpolarized water injected into the NMR are comixed and maintained at the target temperature –the protein pre- and post-injection solutions as regulated in-situ by the NMR console's VT system, and the hyperpolarized water as controlled by a heating-tape-based system. As long as the water is injected above ≈35 ˚C (the temperature at which it arrives from the Hypersense/Arduino system), this pre-calibrated approach should give a solution whose temperature matches that of the protein and hence is constant through space and time.

**Figure S1.** Examining the reliability of HyperW's 2D NMR thermal stability experiments at 50 ˚C. (A, B) Central peak positions observed along the $^1$H (F2) dimension of a 50 ˚C HyperW 2D acquisition, for residue A39 and for the indole W36 peak of the drkN SH3 (unfolded-form) protein –both of which relatively isolated (Fig. 7) and hence can be followed by 1D $^1$H projections. The temperature-dependence of these peaks was independently measured by 2D HMQC NMR, leading to the actual temperatures indicated on the left-hand axes. Shown for completion (dashed lines) are the chemical shifts measured in the 2D HyperW HMQC spectrum for these residues, leading to ≈49.25±0.25 ˚C as the representative temperature of this experiment. (C) 1D traces leading to the results illustrated in panel (B), illustrating both the maxima but also the line shape changes undergone by the W36 indole (U) peak for two post-injection times. The dashed lines indicate the chemical shifts measured for the same peak by conventional 2D HMQC NMR at 47 and 50 ˚C. Notice as well the significant and clearly larger enhancement shown by the folded (F) signal of this residue, as well as its ensuing line narrowing with time.

Still, driven by the importance that temperatures have in defining folded/unfolded protein ratios, particularly for drkN SH3's 50 ˚C dissolution experiments, ancillary analyses were made. Panels A and B in Figure S1 summarize these, by plotting the central peak positions observed along the $^{1}$H (F2) dimension of a 50 ˚C HyperW 2D acquisition, for two residues of the drkN SH3 protein that are both relatively isolated (and hence can be followed in the 1D projection) and show a relatively strong temperature dependence in their $^{1}$H chemical shifts. Independent measurements collected by conventional HMQC NMR in the 47-50 ˚C range, allowed us to translate these HyperW $^{1}$H peak positions into sample temperatures as a function of post-dissolution time. These values are also shown in the left-hand axes of these figure's panels, and paint coinciding pictures regarding the system's stabilization following the injection. According to these, there is a certain drop in the sample temperature immediately upon injection of the pre-heated water (to ≈48 ˚C), but a nearly perfect thermal stabilization at the targeted 50 ˚C temperature ca. 15 s past the injection. These examinations can be extended by analyzing the central peak positions displayed by the residues along the $^{1}$H (F2) dimension, following full 2D processing of the HyperW data. These Fourier-averaged positions should reveal the most representative temperature (and peak intensities) reflected in the 2D HMQC experiments, and they end up being between 49 and 49.5 ˚C (dashed lines in Figures S1A, S1B). Further insight into how samples in 2D HyperW HMQC experiments reach their final temperatures can be gathered from the line shape changes exhibited by the traces illustrated in Figure S1C, which focus on the unfolded SH3 W36 indole resonance –one of the thermally-sensitive peaks that were examined. When analyzed in the ($t_1$,F2)-domain for two different post-dissolution times $t_1$ and compared with traces collected for fully-thermally-equilibrated, $H_2O$-dissolved samples, one can notice that peaks are broader at the beginning of the HyperW series (with a maximum at the equivalent of ca. 48˚C), most likely reflecting a thermal distribution within the NMR tube. Subsequently, peaks sharpen up and coincide with the unfolded chemical shift recorded on a thermally polarized, thermally stabilized sample at 50 ˚C. This sharpening leads to dominating peak positions and intensities corresponding to ca. 49.25±0.25 ˚C, when Fourier processing the full 2D HyperW HMQC interferogram. *In addition, notice how the 1D 17s post-injection trace highlights the orders-of-magnitude enhancements that the folded resonance in this residue gains from the hyperpolarized water over its unfolded counterpart –which of course is one of the paper's main findings.*

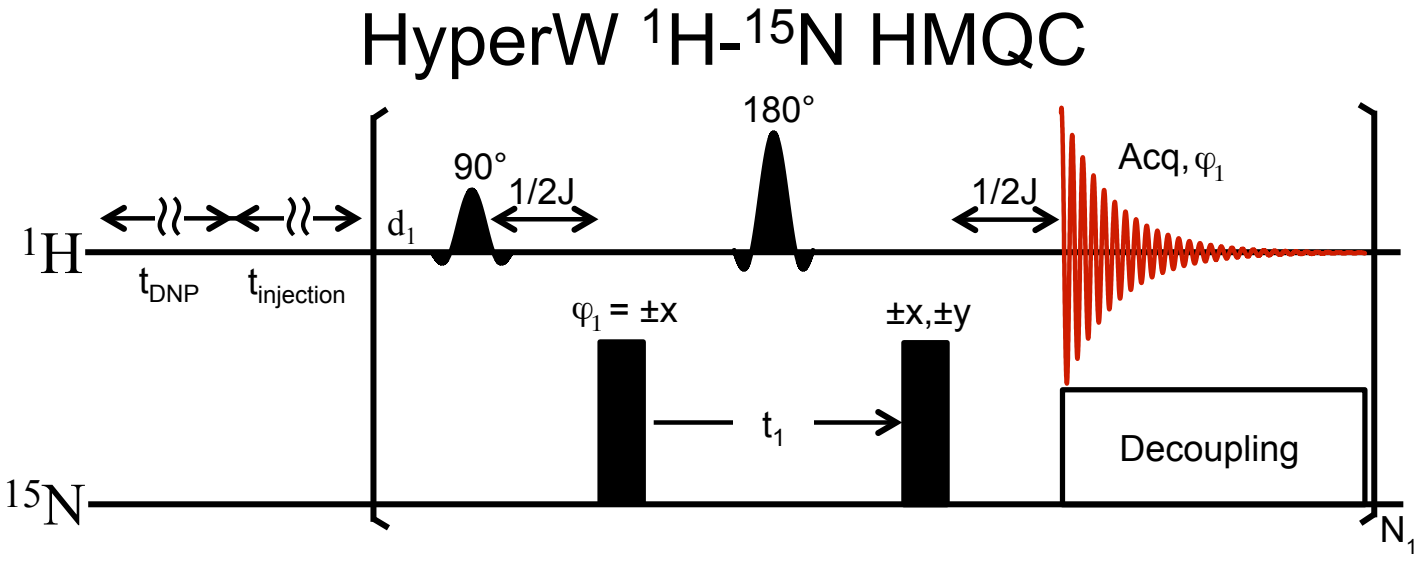

**Figure S2.** Pulse sequence for the 2D HyperW $^{1}$H-$^{15}$N HMQC used in this study. Full bars represent 90° hard pulses, and shapes represent amide-selective 90° and 180° pulses. The recycle delay $d_1$ was typically set to 0.037-0.1 s; water polarization was achieved during $t_{DNP}$ ≈120-180 min, and the subsequent dissolution and injection of hyperpolarized water occurred during $t_{injection}$ ≈2-3 s. Selective excitation of the amide protons was achieved using a PC9 polychromatic pulse,[5] refocusing with a REBURP pulse[6] centered at 8.5 ppm with a 3.0 ppm bandwidth, and typically $N_1$=128 increments were collected. The sequence employed the indicated phase-cycling of the $^{15}$N excitation and storage pulses, to reduce the water background and to deliver by a hypercomplex acquisition purely absorptive lineshapes.[3-4] Decoupling on the $^{15}$N channel was done using GARP modulation[7] during the acquisition.

The two dimensional spectra were acquired using a 2D HyperW $^{1}H$-$^{15}N$ HMQC sequence, similar to what was used in previous work (Fig. S2).[1-2] It is a 2D HMQC-based sequence, using a solute-specific excitation approach.[3-4] The amide downfield region is excited using a selective 90° pulse in order to maximize the use of the hyperpolarized exchanged sites while minimizing water depolarization.

**PhoA$^{(350-471)}$: Per protio and SSP results.** It is of interest to compare the HyperW HMQC spectrum (Fig. 1, red) not only to a thermal equilibrium spectrum measured on the same post-dissolution sample (Fig. 1, blue), but also against a thermal spectrum measured in an 82.5% $H_2O$ buffer under otherwise same conditions. A comparison between these data (Fig. S3) and Fig. 1 reveals that indeed with 82.5% water, one could observe peaks that are broadened beyond detection in the thermal post-dissolution sample, albeit with very poor sensitivity. Peaks are observed with a better sensitivity in the HyperW spectrum, thanks to a nearly ~500× enhancement.

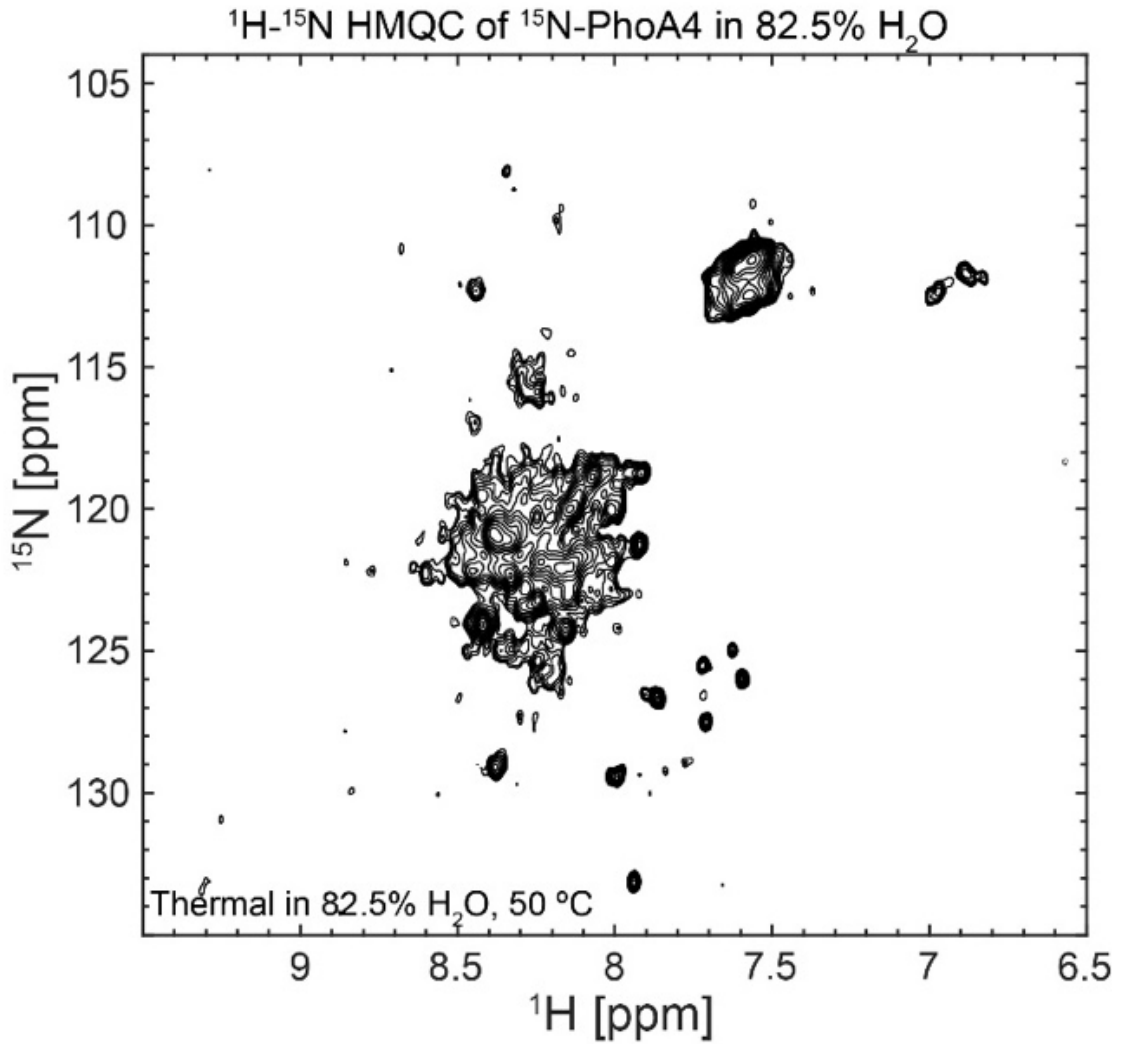

**Figure S3.** Conventional $^{1}H$-$^{15}N$ HMQC spectrum for 0.125 mM $^{15}N$-PhoA4 dissolved in 82.5% $H_2O$ buffer (50 mM HEPES, pH 7.5, 50 mM KCl). The spectrum was recorded at 50 °C using 64 hypercomplex $t_1$ increments covering indirect- and direct-domain spectral widths of 6009.6 and 1825.8 Hz. Additional experimental parameters: 14.1 T Prodigy®-equipped Bruker Avance III NMR spectrometer; total experimental time of 42 min 56 sec (16 scans recorded per $t_1$ increment, acquisition time of 213.0 ms, repetition delay of 1 s). The peaks in the bottom part of the spectrum arise due to slight protein degradation.

Unlike what had been previously reported for α-synuclein, the sensitivity enhancements evidenced by PhoA's HyperW HMQC, do not appear to correlate with the electrostatic charges in the protein sequence (Fig. 2). To explore potential correlations between the enhancements and the SSP values, Supporting Figure S4 compares both individual enhancements vs SSP scores, as well as the running-average enhancements for every three consecutive residues against the SSP average score of the same three residues. A modest correlation appears to emerge in the latter, but it is hard to ascertain.

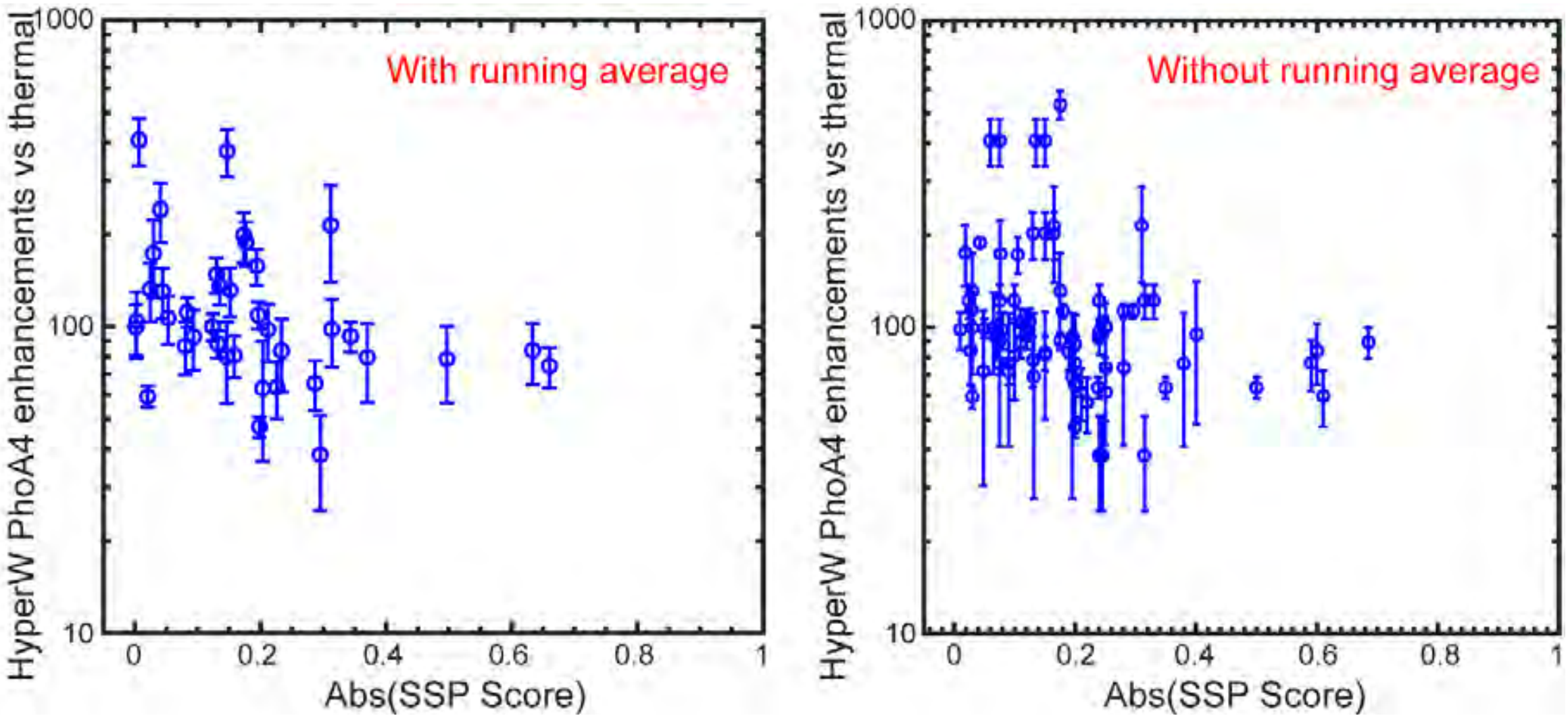

**Figure S4.** Correlations between the HyperW HMQC sensitivity enhancements calculated for resolved residues in the $^{15}N$-labeled PhoA4 protein fragment, against SSP scores (absolute values) given in the literature[20] based on NMR $^{13}C_\alpha$ and $^{13}C_\beta$ chemical shifts. (A) Average enhancements calculated for every three consecutive residues are compared against the absolute value of the average SSP score for the same three consecutive residues. (B) Idem but without the running average.

**Variable temperature ZZ-exchange NMR on drkN SH3.** ZZ-exchange is a kinetic experiment based on a 2D NMR $^1$H-X chemical shift correlation,[9-10] with the addition of a mixing delay $T$ during which the magnetization is stored along the z-axis while dynamics take place. The SH3 domain exists in two exchanging states (U and F), such that a given nucleus resonates at a frequency $\omega_U$ in state U and $\omega_F$ in state F. A 2D ZZ-exchange spectrum for this system (Fig. S5A) will thus contain two diagonal-peaks at U($\omega_U^{13C}$, $\omega_U^{1H}$) and N($\omega_F^{13C}$, $\omega_F^{1H}$) in the (F1, F2) frequency dimensions, and two cross-peaks at C1($\omega_F^{13C}$, $\omega_U^{1H}$) and C2($\omega_U^{13C}$, $\omega_F^{1H}$), due to the U⇋F exchange occurring during the mixing time. In order to obtain kinetic information, a series of ZZ-exchange spectra are recorded with a range of mixing delays. The dependence of the peak intensities on the mixing delay are then analyzed and fit to a kinetic exchange model (Fig. S5B):[11]

(S1) $$\frac{I_U(T)}{I_U(0)} = A_U \cdot \frac{-(\lambda_2 - a_{22})\cdot e^{-\lambda_1 T} + (\lambda_1 - a_{22})\cdot e^{-\lambda_2 T}}{\lambda_1 - \lambda_2}$$

(S2) $$\frac{I_N(T)}{I_N(0)} = A_F \cdot \frac{-(\lambda_2 - a_{11})\cdot e^{-\lambda_1 T} + (\lambda_1 - a_{11})\cdot e^{-\lambda_2 T}}{\lambda_1 - \lambda_2}$$

(S3) $$\frac{I_{C1}(T)}{I_U(0)} = A_F \cdot \frac{a_{12}\cdot e^{-\lambda_1 T} - a_{12}\cdot e^{-\lambda_2 T}}{\lambda_1 - \lambda_2}$$

(S4) $$\frac{I_{C2}(T)}{I_N(0)} = A_U \cdot \frac{a_{21}\cdot e^{-\lambda_1 T} - a_{21}\cdot e^{-\lambda_2 T}}{\lambda_1 - \lambda_2} ,$$

where $\lambda_{1,2} = \frac{1}{2} \cdot \{(a_{11} + a_{22}) \pm [(a_{11} - a_{22})^2 + 4k_{F\to U}k_{U\to F}]^{1/2}\}$, $a_{11} = R_1^F + k_{U\to F}$, $a_{12} = -k_{F\to U}$, $a_{22} = R_1^U + k_{F\to U}$ and $a_{21} = -k_{U\to F}$. $R_1^U$ and $R_1^F$ are the longitudinal relaxation rate constants of magnetization in sites U and F, respectively, and $I_U(0)$ and $I_N(0)$ are the peak intensities in the unfolded and folded states, respectively, at $T = 0$. The factors $A_F$ and $A_U$ represent efficiency of coherence transfer after the mixing period $T$, and were determined as described previously.[12] The simultaneous fits to the data yield the first-order rate constants $k_{F\to U} = 31.0 \pm 4$ $s^{-1}$ and $k_{U\to F} = 1.9 \pm 0.4$ $s^{-1}$. The populations at 50 °C are therefore: $p_U = 94.3\%$ and $p_F = 5.7\%$; these ZZ-exchange results (Fig. S5), which take into account compensation for differences in the relaxation rates ($R_1$ and $R_2$) of the peaks. The folded state population of only 5.7% at 50 °C, is to be contrasted to the 55% observed at 27 °C. Populations and exchange rates at 37 °C were also calculated using a $^1H$-$^{15}N$ version of the ZZ-exchange experiment, as $k_{F\to U,37C} = 7.9 \pm 2.8$ $s^{-1}$, $k_{U\to F,37C} = 8.7 \pm 2.9$ $s^{-1}$, $p_{U,37C} = 48\%$ and $p_{F,37C} = 52\%$. Table S1 summarizes these kinetic and population values, as derived by these measurements on SH3 at the three temperatures that we explored.

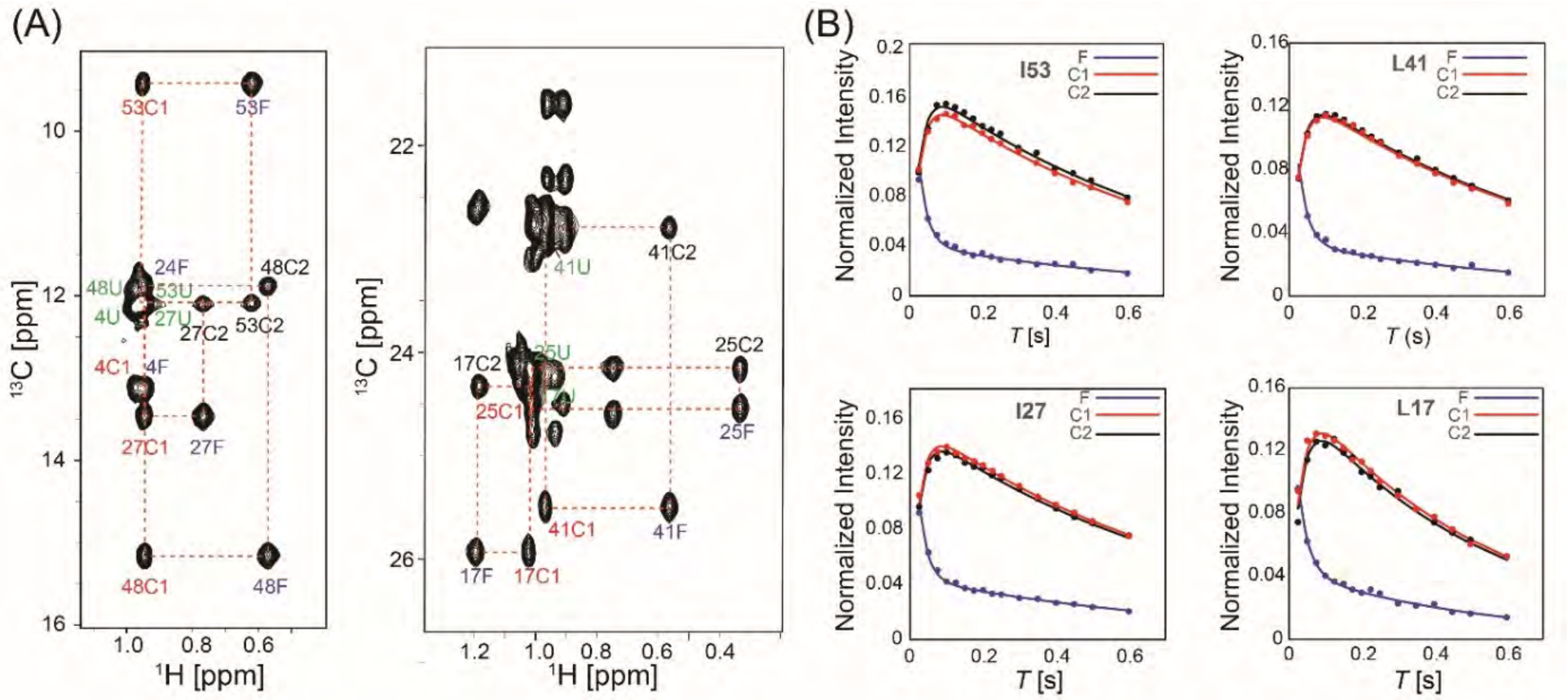

**Figure S5.** ZZ-exchange experiment to extract kinetic information on the drkN-SH3 domain U⇋F exchange. (A) Selected regions from one ZZ-exchange spectrum (corresponding to a mixing period of $T = 0.05$ s), measured on an 800 MHz spectrometer, equipped with a cryo-cooled HCN probe. Indirect- and direct-domain spectral widths of 12019.2 and 4022 Hz were covered, using 82 hypercomplex $t_1$ increments and STATES acquisition.[8] $N_S = 16$ scans were collected using a 3 s acquisition time, and a relaxation delay of 1.5 s. Total experimental time was ~ 3 hrs 20 min for each different mixing time $T$. The assigned residues are denoted by their residue number, and the peaks are labeled as: U – unfolded diagonal-peak, F – native (folded) diagonal-peak, C1 and C2 – cross-peaks. (B) Normalized peak intensities as a function of mixing times for selected residues, for the diagonal-peak and the C1 and C2 cross peaks. The data points are fitted to Eqs. (S2), (S3) and (S4) to extract kinetic information.

***Table S1.*** *Kinetic parameters for the U⇋F process of drkN SH3 domain, as derived from ZZ-exchange measurements for 50 °C, 100% $D_2O$; and 37 °C, 90% $H_2O$; and from HMQC peak intensity ratios at 27 °C, 90% $H_2O$.*

| ***Temperature (°C)*** | ***Relative population of the folded state ($p_F$, %)*** | ***Relative population of the unfolded state ($p_U$, %)*** | ***U⇋F exchange rate $k = k_{F\to U} + k_{U\to F}$ ($s^{-1}$)*** | ***$k_{F\to U}$ ($s^{-1}$)*** | ***$k_{U\to F}$ ($s^{-1}$)*** |
|---|---|---|---|---|---|
| 50 | 6 ± 10 | 94 ± 10 | 33 ± 4 | 31 ± 4 | 1.9 ± 0.4 |
| 37 | 52 ± 10 | 48 ± 10 | 17 ± 3 | 8 ± 3 | 9 ± 3 |
| 27 | 60 ± 20 | 40 ± 20 | | | |

**Methyl-TROSY NMR experiments on drkN SH3.** The relative populations in Table S1 were derived from ZZ-exchange experiments performed on mostly deuterated (90% $D_2O$) solutions. $D_2O$ as a solvent, however, has been reported to stabilize certain protein structures when compared to $H_2O$, and to affect protein folding-unfolding kinetics.[21,22] The dDNP enhancements reported for drkN SH3, however, are done by comparing HyperW results arising from deuterated solutions, against thermal results arising from mostly protonated ones. It follows that in order to properly quantify the signal enhancement upon hyperpolarization, one must also take into account the potential differences in the populations observed for the folded states in $D_2O$ (the solvent used in the HyperW measurements) vs. in $H_2O$ (the solvent used for the thermal equilibrium measurements). As solvent exchanges prevent us from measuring the populations of drkN SH3's folded and unfolded forms by relying on the amide group resonances using $D_2O$ as solvent, a methyl labeled ($^{13}CH_3$-ILVM, $^{2}H$) drkN SH3 protein was expressed, and the populations of these two forms were quantified by integrating the peak intensities of ten residues in the folded and unfolded states using methyl-TROSY $^{1}H$-$^{13}C$.[23] These experiments were carried out at 50 ˚C in both 90% $H_2O$ and in 100% $D_2O$. The ensuing results are summarized in Table S2. As can be appreciated from these results, the populations of the folded states at 50 °C in $D_2O$ were indeed higher than those in $H_2O$: 5.7 ± 0.6%.vs 4.0 ± 0.7%. At the same time, the deuterated solvent methyl-TROSY results were in excellent agreement with the ZZ-exchange measurements. These methyl-TROSY-derived populations were used in the simulations described throughout the Supporting and the Main texts; these populations were also used to rescale the kinetic rates in Table S1, as appropriate.

***Table S2.*** *Folded-state populations extracted for various methyl residues of drkN SH3, as derived from methyl-TROSY measurements performed at 50 °C and 90/10% mixtures of mostly per-deutero and per-protio aqueous solutions. Data were recorded at 800 MHz using a TCI Cryoprobe®.*

| **drkN SH3 Residue** | **% - Folded state population** | |
|---|---|---|
| | Solvent – $D_2O$ | Solvent – $H_2O$ |
| Ile4 $\delta_1$ | 5.9 | 3.8 |
| Leu17 $\delta$ | 5.8 | 4.2 |
| Leu25$\delta$ | 5.8 | 3.3 |
| Ile27 $\delta$ | 4.2 | 2.9 |
| Leu41 $\delta_1$ | 5.3 | 4.0 |
| Leu41 $\delta_2$ | 5.7 | 3.4 |
| Ile48 $\delta_1$ | 6.3 | 5.6 |
| Ile53 $\delta_1$ | 6.8 | 3.2 |
| Leu30 $\delta$ | 5.6 | 3.8 |
| Leu50 $\delta$ | 5.9 | 4.9 |

**Supporting Figure S6** compares CLEANEX-PM measurements at 37 °C, with the HyperW enhancements observed for the folded and unfolded residues of the drkN-SH3 domain.

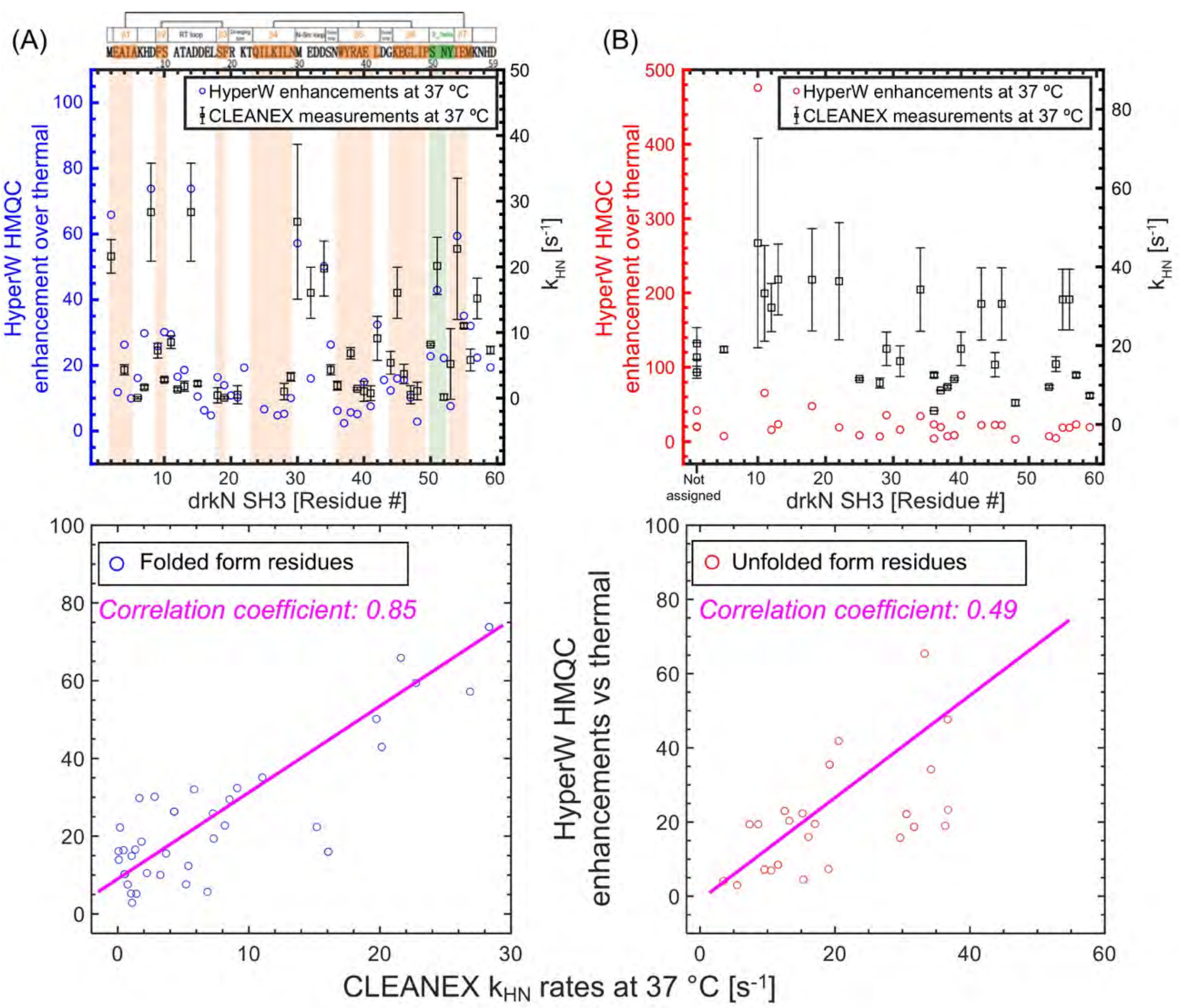

**Figure S6.** Comparison of HyperW enhancements and CLEANEX-derived exchange rates for the folded (left) and unfolded (right) states of drkN SH3 domain at 37 °C. (A) Comparison of amide proton exchange rates $k_{FW}$ arising for different drkN SH3 residues in the folded state as extracted from CLEANEX-PM experiments[13] at 14.1 T and 37 °C (black squares), with the corresponding HyperW HMQC sensitivity enhancements at 37 °C (blue circles, taken from Fig. 8B). Orange and green shaded areas are drawn in the bottom panel for regions which correspond to the secondary structure elements depicted in the top panel (as in Figs. 8A and 8B). The linear correlation coefficient (bottom), between the HyperW enhancements and CLEANEX-PM exchange rates for the folded state at 37 °C is 0.85. (B) Comparison of amide proton exchange rates $k_{UW}$ arising for different drkN SH3 residues in the unfolded state as extracted from CLEANEX-PM experiments[13] at 14.1 T and 37 °C (black squares), with the corresponding HyperW HMQC sensitivity enhancements at 37 °C (red circles, taken from Fig. 8C). The correlation coefficient (again calculated for the data in a linear plot) between the HyperW enhancements and CLEANEX exchange rates for the unfolded state at 37 °C (bottom) was 0.49.

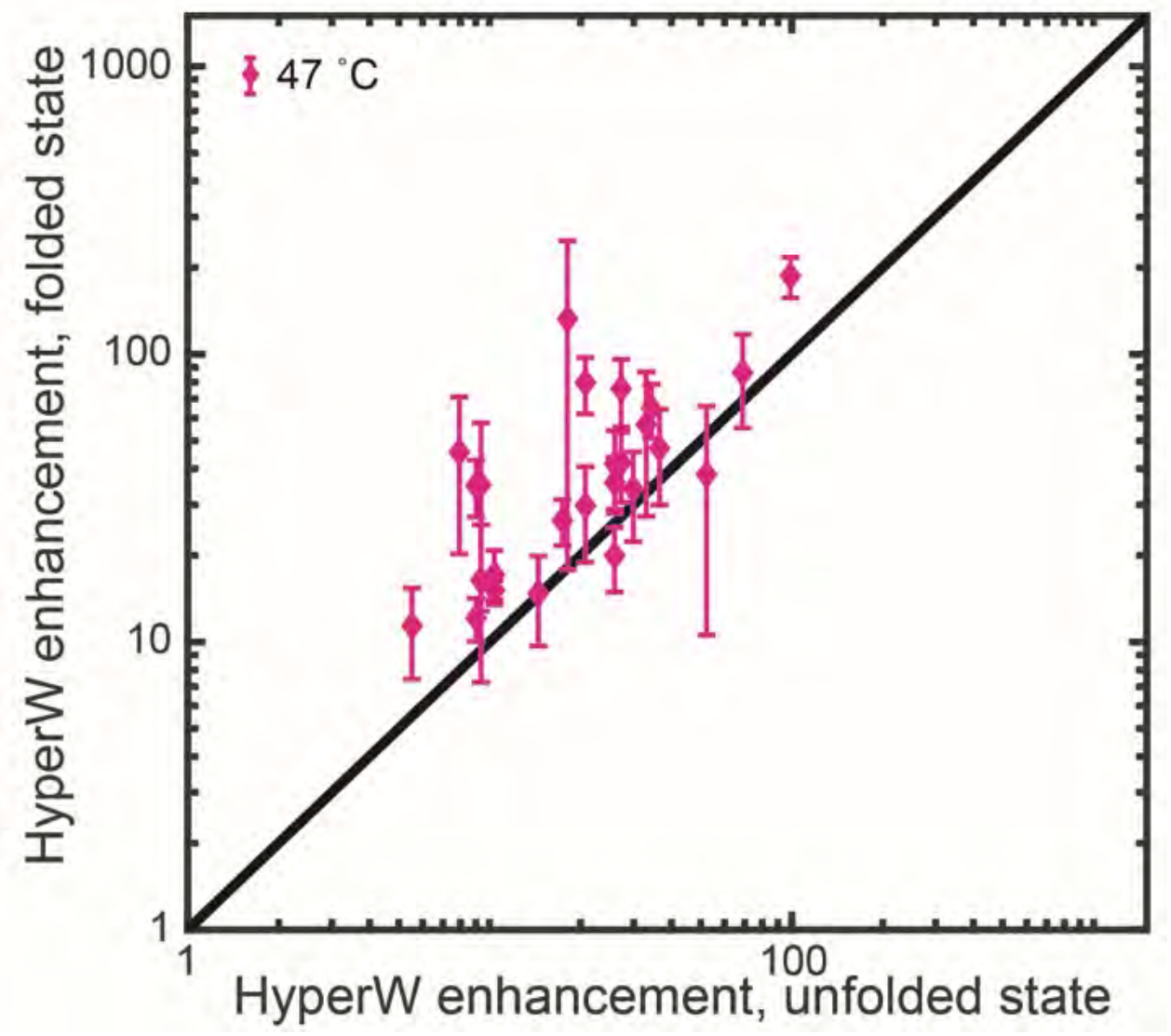

**Figure S7.** Same as Figure 8E in the main text, but assuming that the HyperW injection temperature had been misscalibrated and actually took place at 47 ˚C.

**Supporting Figure S7** re-examines drkN SH3 HyperW's enhancements measured at 50 °C, assuming that post-injection temperatures were not as believed but instead lower by 3 ˚C –a difference that is still compatible with the peak positions recorded in the HMQC NMR spectra. Indeed, although Fig. S1 attests to the good thermal reliability of our setup, the linewidths of the HyperW NMR data yield a certain uncertainty in the temperature, which is bound by a lower limit of 47 ˚C. This plot is a recalculation of the enhancement data presented in Figure 8E, but with enhancements renormalized according to thermally polarized reference spectra measured at 47 ˚C. As evidenced by this plot, this still leads to a picture where folded-residue peaks are more enhanced by the hyperpolarized solvent than their unfolded-state counterparts.

Table S3 summarizes the enhancements observed for the various folded and unfolded drkN SH3 residues at 50 ˚C, taking into account multiple dissolutions and the population considerations in Table S2. Comments indicate why the corresponding residues were not utilized in the paper's discourse/conclusions.

***Table S3.*** *Average enhancements ± deviation obtained from three separate hyperpolarized water injections for folded and unfolded residues of the drkN SH3 domain at 50˚C.*

| *Folded/Unfolded* | *Residue* | *Enhancement±deviation* | *Comments* |
|---|---|---|---|
| F | E2 | 170 ± 50 | |
| F | A3 | 200 ± 60 | |
| F | I4 | 70 ±10 | |
| F | A5 | 100 ± 20 | |
| F | K6 | 120 ± 20 | |
| F | H7 | 30 ± 40 | Doesn't appear in dissolution #1, and very weak in dissolution #2 – overlap with sidechains |
| F | D8 | 90 ± 40 | |
| F | F9 | 120 ± 20 | |
| F | S10 | 230 ± 20 | |
| F | A11 | 50 ± 30 | |
| F | T12 | 160 ± 40 | |
| F | A13 | 210 ± 90 | |
| F | D14 | 90 ± 40 | |
| F | D15 | 100 ± 20 | |
| F | E16 | 49 ± 8 | |
| F | L17 | 58 ± 5 | |
| F | S18 | 340 ± 20 | |
| F | F19 | 240 ± 70 | |
| F | R20 | 75 ± 8 | |
| F | K21 | 75 ± 1 | |
| F | T22 | 200 ± 40 | |
| F | Q22 | | |
| F | I23 | | |
| F | L25 | 50 ± 20 | |
| F | K26 | | |
| F | I27 | 58 ± 5 | |
| F | L28 | 42 ± 4 | |
| F | N29 | 96 ± 9 | |
| F | M30 | 130 ± 50 | |
| F | E31 | Not identified | |
| F | D32 | 23 ± 3 | |
| F | D33 | Not identified | |
| F | S34 | 180 ± 200 | Heavy overlap |
| F | N35 | 70 ± 10 | |
| F | W36 | Not identified | |
| F | W36_INDOLE | 70 ± 30 | |
| F | Y37 | 24 ± 2 | |
| F | R38 | 57 ± 6 | |
| F | A39 | 90 ± 50 | |
| F | E40 | 100 ± 10 | |
| F | L41 | 80 ± 20 | |
| F | D42 | 300 ± 40 | |
| F | G43 | 130 ± 20 | |
| F | K44 | 39.1 ± 0.8 | |
| F | E45 | 23 ± 3 | |
| F | G46 | 120 ± 10 | |
| F | L47 | 75.4 ± 0.9 | |
| F | I48 | 29 ± 6 | |
| F | P49 | Not identified | |
| F | S50 | 31 ± 2 | |
| F | N51 | 140 ± 20 | |
| F | Y52 | 110 ± 10 | |
| F | I53 | Not identified | |
| F | E54 | 31.0 ± 0.8 | |
| F | M55 | 190 ± 20 | |
| F | K56 | 65 ± 10 | |
| F | N57 | 30 ± 2 | |
| F | H58 | Not identified | |
| F | D59 | 19 ± 4 | |
| U | E2 | Not identified | |
| U | A3 | Not identified | |
| U | I4 | Not identified | |
| U | A5 | 55 ± 20 | |
| U | K6 | Not identified | |
| U | H7 | Not identified | |
| U | D8 | Not identified | |
| U | F9 | Not identified | |
| U | S10 | 60 ± 50 | Hard to identify |
| U | A11 | 100 ± 90 | Overlap |
| U | T12 | 60 ± 20 | |
| U | A13 | 75 ± 70 | Overlap |
| U | D14 | Not identified | |
| U | D15 | Not identified | |
| U | E16 | Not identified | |
| U | L17 | Not identified | |
| U | S18 | 200 ± 200 | Heavy overlap |
| U | F19 | Not identified | |
| U | R20 | Not identified | |
| U | K21 | Not identified | |
| U | T22 | 85 ± 70 | Ambiguous assignment |
| U | Q22 | Not identified | |
| U | I23 | Not identified | |
| U | L25 | 18.7 ± 0.5 | |
| U | K26 | Not identified | |
| U | I27 | Not identified | |
| U | L28 | 18 ± 2 | |
| U | N29 | 53 ± 2 | |
| U | M30 | Not identified | |
| U | E31 | 33 ± 2 | |
| U | D32 | Not identified | |
| U | D33 | Not identified | |
| U | S34 | 40 ± 10 | |
| U | N35 | Not identified | |
| U | W36 | 36 ± 1 | |
| U | W36_INDOLE | 13.75 ± 0.04 | |
| U | Y37 | 19 ± 2 | |
| U | R38 | 16 ± 2 | |
| U | A39 | 18.7 ± 0.5 | |
| U | E40 | 53 ± 2 | |
| U | L41 | Not identified | |
| U | D42 | Not identified | |
| U | G43 | 39 ± 6 | |
| U | K44 | Not identified | |
| U | E45 | 44 ± 2 | |
| U | G46 | 39 ± 6 | |
| U | L47 | Not identified | |
| U | I48 | 9 ± 2 | |
| U | P49 | Not identified | |
| U | S50 | Not identified | |
| U | N51 | Not identified | |
| U | Y52 | Not identified | |
| U | I53 | 16 ± 2 | |
| U | E54 | 16 ± 6 | |
| U | M55 | 31 ± 8 | |
| U | K56 | 31 ± 8 | |
| U | N57 | 36 ± 1 | |
| U | H58 | Not identified | |
| U | D59 | 27 ± 3 | |
| U | Overlap of multiple sites | 35 ± 1 | Broad resonance centered at 8ppm $^{1}$H. 120.6ppm $^{15}$N |
| U | Overlap of multiple sites | 46 ± 2 | Broad resonance centered at 8.2ppm $^{1}$H. 120.5ppm $^{15}$N |
| U | Overlap of multiple sites | 58 ± 1 | Broad resonance centered at 8.35ppm $^{1}$H. 118.5ppm $^{15}$N |

**$^{1}$H-$^{15}$N HyperW HMQC for a three-site exchanging system incorporating cross-relaxation: Theoretical Considerations.** Site-specific amide-water exchange rates lead to heterogeneities in the HyperW enhancement. However, the exchange rates in the folded state were expected to be slower relative to the unfolded state, due to protection factors and hydrogen bonds. Scheme 1 suggests that an additional magnetization transfer from an enhanced unfolded state residue to the same residue in the folded state can explain its observed sensitivity enhancements. Biases in the hyperpolarization of folded and unfolded residues could also arise from the different cross-relaxation behavior of these systems. To estimate how the HyperW signal enhancements will be affected by these exchanges, we computed the water and amide magnetizations for each conformation $\langle H_2O\rangle_z$, $\langle H_N^U\rangle_z$ , $\langle H_N^F\rangle_z$ expected to arise in a process characterized by a forward reaction rate (proton transfers from $H_2O$ to $H_N$) $k_{WU}$, $k_{WF}$; and a backward reaction rate $k_{UW}$ , $k_{FW}$. These exchange rates are in fact related to each other by the water and protein molar fraction ratios $X$:

$$k_{UW} = \frac{X_{H_2O}}{X_{H_N{}^U}} \cdot k_{WU} \quad , \quad k_{FW} = \frac{X_{H_2O}}{X_{H_N{}^F}} \cdot k_{WF} \qquad \text{(S5).}$$

A model based on the McConnell-Solomon equations[14-16] was implemented within a home-written Matlab® (The Mathworks Inc.) code that involved numerical solution of the system of differential equations for different proton reservoirs, including chemical exchange and cross-relaxation between amide and aliphatic proton pools as well as with protons in the hyperpolarized $H_2O$ pool. This leads to the system of 7 differential equations:

$$\frac{d}{dt}\begin{pmatrix} \langle H_N{}^F\rangle_z(t) - \langle H_N{}^F\rangle_z(eq) \\ \langle H_N{}^U\rangle_z(t) - \langle H_N{}^U\rangle_z(eq) \\ \langle H_C{}^F\rangle_z(t) - \langle H_C{}^F\rangle_z(eq) \\ \langle H_C{}^U\rangle_z(t) - \langle H_C{}^U\rangle_z(eq) \\ \langle H_X{}^F\rangle_z(t) - \langle H_X{}^F\rangle_z(eq) \\ \langle H_X{}^U\rangle_z(t) - \langle H_X{}^U\rangle_z(eq) \\ \langle H_2O\rangle_z(t) - \langle H_2O\rangle_z(eq) \end{pmatrix} =$$

$$\begin{pmatrix} -r_F & k_{UF} & \sigma_F & 0 & \sigma_{XF} & 0 & k_{WF}+\sigma_{WF} \\ k_{FU} & -r_U & 0 & \sigma_{CU} & o & \sigma_{XU} & k_{WU}+\sigma_{WU} \\ \sigma_F & 0 & -R_1^{H_CF} & k_{UF} & 0 & 0 & \sigma_{WF} \\ 0 & \sigma_U & k_{FU} & -R_1^{H_CU} & 0 & 0 & \sigma_{WU} \\ \sigma_{XF} & 0 & \sigma_{XF} & 0 & -R_1^{H_XF} & k_{UF} & k_{WX}+\sigma_{WFX} \\ 0 & \sigma_{XU} & 0 & \sigma_{XU} & k_{FU} & -R_1^{H_XU} & k_{WX}+\sigma_{WUX} \\ k_{FW}+\sigma_{FW} & k_{UW}+\sigma_{UW} & 0 & 0 & k_{XW}+\sigma_{WFX} & k_{XW}+\sigma_{WUX} & -r_W \end{pmatrix}\begin{pmatrix} \langle H_N{}^F\rangle_z(t) - \langle H_N{}^F\rangle_z(eq) \\ \langle H_N{}^U\rangle_z(t) - \langle H_N{}^U\rangle_z(eq) \\ \langle H_C{}^F\rangle_z(t) - \langle H_C{}^F\rangle_z(eq) \\ \langle H_C{}^U\rangle_z(t) - \langle H_C{}^U\rangle_z(eq) \\ \langle H_X{}^F\rangle_z(t) - \langle H_X{}^F\rangle_z(eq) \\ \langle H_X{}^U\rangle_z(t) - \langle H_X{}^U\rangle_z(eq) \\ \langle H_2O\rangle_z(t) - \langle H_2O\rangle_z(eq) \end{pmatrix}$$

(S6)

The relaxation matrix used in this model was generated using the Bloch-Redfield-Wangsness theory on a reduced spin system,[17] in combination with SpinDynamica.[18] Five spins were included to account for the polypeptide backbone and sidechain ($H_N$, $N_H$, two aliphatic sidechain protons $H_c^1$ and $H_c^2$, one labile sidechain protons $H_X$ ) and a reduced relaxation matrix with only longitudinal terms for each spin present was utilized. A model-free approach with order parameters for each interaction was adopted,[19] with spectral densities given by the general form:

$$J_i(\omega) = \frac{2}{5}S_i^2\frac{\tau_c}{1+(\omega\tau_c)^2} \quad \text{(S7)}$$

The final relaxation matrix includes two spin order longitudinal terms of folded and unfolded conformations for the amide protons, two corresponding terms for the aliphatic protons, one for the sidechain labile proton, and one for the external water proton. Cross-relaxation between the amide and aliphatic spin pools were assumed to differ for the folded and unfolded states, given in each case by:

$$\sigma = \frac{1}{10}\delta_{HH}^2[J(2\omega_H) - J(0)] \quad \text{(S8)}$$

where $\delta_{HH} = (\mu_0\gamma_H^2 h)/(8\pi d_{HH}^3)$. Diagonal elements in the relaxation matrix are given by $r_F = k_{FW} + k_{FU} + R_1^{H_NF} + \frac{1}{T_1^F}$, $r_U = k_{UW} + k_{UF} + R_1^{H_NU} + \frac{1}{T_1^U}$, $R_1^{H_CF}$, $R_1^{H_CU}$ and $r_W = \frac{1}{T_1^W} + k_{WU} + k_{WF}$ . The rates $k_{FU}$ and $k_{UF}$ represent the exchange rates between the folded and unfolded states (see Scheme 1). $R_1^{H_NF}, R_1^{H_NU}, R_1^{H_CF}$ and $R_1^{H_CU}$ are the corresponding auto-relaxation rates of the amide and aliphatic protons in the folded and unfolded states, including dipolar interactions between the $^{15}N$ and $H_N$, $H_N$ and $H_C$, as well as between pairs $H_C$-$H_C$ of aliphatic protons. Based on this, the rates are then given by

$$\begin{aligned} R_1^{H_N} &= \frac{1}{10}\left(\delta_{H_N N}^2\left[J_{H_N N}(\omega_N-\omega_N)+3J_{H_N N}(\omega_N)+6\,J_{H_N N}(\omega_N+\omega_H)\right]\right.\\ &\qquad \left.+\,2\times\delta_{H_N H_C}^2\left[J_2(0)+3J_2(\omega_H)+6\,J_2(2\omega_H)\right]\right)\\ R_1^{H_C} &= \frac{1}{10}\left(\delta_{H_C H_C}^2\left[J_2(0)+3J_2(\omega_H)+6\,J_2(2\omega_H)\right]\right.\\ &\qquad \left.+\delta_{H_C H_N}^2\left[J_2(0)+3J_2(\omega_H)+6\,J_2(2\omega_H)\right]\right) \end{aligned} \tag{S9}$$

Additional intrinsic relaxation rates $1/T_1^{U,F}$ were also added to the relaxation terms of each amide proton, in a search for an additional *ad hoc* parameter that might potentially explain drkN SH3's anomalous HyperW behavior. Order parameters and internuclear distances were chosen from the literature for Ubiquitin at room temperature, which is a fair assumption based on the very similar molecular weights of Ubiquitin and drkN-SH3 domain. The $R$ and $\sigma$ rates will mostly depend on the internuclear amide/aliphatic distance $d_{HH}$ (kept constant at 2.3 Å for the folded and unfolded states for simplicity) and on the internuclear correlation time $\tau_c$ , which was taken to be 3.4 ns for the folded state and 0.8 ns for the unfolded state of the protein. As purely intramolecular cross-relaxation models failed to predict larger folded than unfolded enhancements unless exchange rates $k_{FW} \geq k_{UF}$ were invoked, Eq. (S6) was modified to enable the presence of intermolecular water-amide proton-proton cross-relaxation. This interaction was incorporated into the simulations in a manner similar to that in Eq. (S8); for simplicity, the same correlation times were assumed to control the intra- and inter-molecular cross-relaxation processes (3.4 ns for the folded, 0.8 ns for the unfolded H—H vectors). $\langle H_2O\rangle_z(eq)$, $\langle H_N\rangle_z(eq)$ and $\langle H_C\rangle_z(eq)$ in Eq. (S6) are the water and protein amide and aliphatic magnetizations at thermal equilibrium. Complementing Eq. (S6)'s time-dependence, the evolution of $\langle H_N\rangle_z(t)$ was artificially set to zero at $t = n \cdot TR$ (where $TR$ is the experimental repetition time) to account for the depletion of protein magnetization arising due to the selective excitation pulses applied. Equation (S6) plus this reset condition were used for analyzing both the HyperW (*Hyp*) and the thermal equilibrium (*TE*) experiments that were carried out, which were recorded on the same samples under identical conditions –apart from their initial water polarization. The initial water magnetization was $\langle H_2O\rangle_z(0) = \varepsilon \cdot \langle H_2O\rangle_z(eq)$, where ε is the enhancement factor over the thermal equilibrium polarization (ε = 200 for the *Hyp* experiment; and ε = 1 for the *TE* experiment). The initial polarization for the amide protons in the protein was assumed to be $\langle H_N\rangle_z(0) = 0$; $\langle H_C\rangle_z(0)$ and $\langle H_C\rangle_z(eq)$ were set equal to unity. For both cases (*Hyp* and *TE*) the equilibrium polarization was scaled according to the concentrations: $\langle H_2O\rangle_z(eq) = \frac{X_{H_2O}}{X_{H_N{}^F}} \equiv X_{WF}$ ; $\left\langle H_N{}^U\right\rangle_z(eq) = \frac{X_{H_N{}^U}}{X_{H_N{}^F}} \equiv X_{UF}$; $\left\langle H_N{}^F\right\rangle_z(eq) = 1$ and same holds for aliphatic spin pools.

In order to translate the magnetizations that will be predicted by these equations into observable signals, we further considered that in the full 2D HyperW $^1H$-$^{15}N$ HMQC experiment these will have to be converted into a $^1H$ coherence that transfers to and from the amide nitrogens:

$$\langle H_N\rangle_z \xrightarrow{pulse\ P} \langle H_N\rangle_x \xrightarrow{J_{HN}} \langle {}^{15}N\rangle_x \xrightarrow{t_1} \langle {}^{15}N\rangle_{x,y}(t_1) \xrightarrow{J_{HN}} \langle H_N\rangle_{x,y}(detect) \quad \text{(S10).}$$

Besides $T_2$–derived losses that for simplicity were ignored, the efficiency of these coherence transfers/encodings will also depend on the inverse $H_N \rightarrow H_2O$ rate constant ($k_{UW}, k_{FW}$): indeed, rapid exchanges of the amide proton with the solvent will preclude an efficient coherence transfer to the $^{15}N$ via J-couplings, and/or will contribute to the dephasing of the MQ state represented by $\langle {}^{15}N\rangle_x$ evolving during $t_1$. This will lead to an overall exponential signal decay, where the duration of the decay period for the $n^{th}$ $t_1$ increment can be expressed as (see Fig. S2):

$$t_{acq}(n) = P90_H + \frac{1}{J_{HN}} + 2 \cdot P90_N + t_1(n) + P180_H \quad \text{(S11).}$$

Accordingly, we express the average signal per scan after a total of $N_1$ increments $t_1$ as:

$$S_U(TR, k_{UW}, k_{FW}) = \frac{1}{N_1}\left(\sum_{n=1}^{N_1} \langle H_N{}^U\rangle_z (nTR, k_{UW}, k_{FW}) \cdot e^{-k_{UW}\cdot t_{acq}(n)}\right) \quad \text{(S12a)}$$

$$S_F(TR, k_{UW}, k_{FW}) = \frac{1}{N_1}\left(\sum_{n=1}^{N_1} \langle H_N{}^F\rangle_z (nTR, k_{UW}, k_{FW}) \cdot e^{-k_{FW}\cdot t_{acq}(n)}\right) \quad \text{(S12b)}$$

where we stress the potential dependence of the amide magnetization on the time *t* that each $t_1(n)$ increment will have associated since the injection of the hyperpolarized solvent. On the basis of all these considerations, the various 3D plots shown in Fig. 10 of the main text were computed.

## Supporting Information References